\documentclass[11pt]{article}
\usepackage[letterpaper,margin=1in]{geometry}
\usepackage{odonnell}

\newcommand{\Recursive}{\mathsf{R}}
\newcommand{\Rec}{\Recursive}
\newcommand{\RE}{\mathsf{RE}}
\newcommand{\Halts}{\textsc{Halt}}
\newcommand{\cHalts}{\overline{\Halts}}
\newcommand{\cHalt}{\cHalts}

\newcommand{\trnorm}[1]{\norm{#1}_{1}}

\title{Computability Theory of Closed Timelike Curves\footnote{An earlier unpublished version of this paper~\cite{ABG16} contained an erroneous proof of the main theorem, \cite[Theorem~10]{ABG16}.  The error, discovered in~\cite{Raa23}, was asserting $\|A\|_{\mathrm{F}} \leq \|A\|_{\mathrm{tr}}$ in \cite[Proposition~8]{ABG16}. The present work provides a correction in \Cref{thm:main-quantum}.}}

\author{
Scott Aaronson\thanks{The University of Texas at Austin. \texttt{aaronson@cs.utexas.edu}.}
 \and
Mohammad Bavarian\thanks{Work completed as a graduate student at MIT. \texttt{mobavarian@gmail.com}.}
 \and
Toby Cubitt\thanks{University College London. \texttt{t.cubitt@ucl.ac.uk}.}
 \and
Sabee Grewal\thanks{The University of Texas at Austin. Supported by the NSF QLCI Award OMA-2016245. \texttt{sabee@cs.utexas.edu}.}
 \and 
Giulio Gueltrini\thanks{Work completed as an undergraduate student at MIT.}
 \and
Ryan O'Donnell\thanks{Carnegie Mellon University. Supported in part by a grant from Google Quantum AI.
\texttt{odonnell@cs.cmu.edu}.}
 \and
Marien Raat\thanks{Freudenthal Institute, Utrecht University, The Netherlands. \texttt{mail@marienraat.nl} }
}

\date{}

\begin{document}

\maketitle

\begin{abstract}
    We study the question of what is computable by Turing machines equipped with time travel into the past; i.e., with Deutschian closed timelike curves (CTCs) having no bound on their width or length.
    An alternative viewpoint is that we study the complexity of finding approximate fixed points of computable Markov chains and quantum channels of countably infinite dimension.
    
    Our main result is that the complexity of these problems is precisely~$\Delta_2$, the class of languages Turing-reducible to the Halting problem.
    Establishing this as an upper bound for qubit-carrying CTCs requires recently developed results in the theory of quantum Markov maps.
\end{abstract}

\section{How would time travel change the theory of computation?} \label{sec:intro}

A closed timelike curve (CTC) is a cycle in the spacetime
manifold that is locally timelike, and that can therefore carry a particle to
its own past.
CTCs can formally appear in solutions to Einstein's
field equations of general relativity (as shown inadvertently by Lanczos~\cite{Lan24} and van~Stockum~\cite{Sto38}, and explicitly by G\"{o}del~\cite{Goe49}).
But whether they in fact exist, or can be created, is an unsolved problem of physics.

One theoretical approach to the problem lies in reasoning about how information and computation would be affected by the existence of CTCs.
This approach was initiated in the work of Deutsch~\cite{Deu91}, who used it to critique the conclusion (put forth, e.g., by Hawking and Ellis~\cite{HE73}) that time-travel should be impossible thanks to the ``grandfather paradox''.
G\"{o}del himself spelled out this seeming contradiction:
\begin{quotation}
    ``\emph{[B]y making a round trip on a rocket ship in a sufficiently wide curve, it is possible in [worlds with CTCs] to travel into any region of the past, present, and future, and back again, exactly as it is possible in other worlds to travel to distant parts of space.
    This state of affairs seems to imply an absurdity. For it enables one, e.g.\ to travel into the near past of those places where he has himself lived. There he would find a person who would be himself at some earlier period of his life. Now he could do something to this person that, by his memory, he knows has not happened to him. This and similar contradictions\dots}''\footnote{Unlike Hawking and Ellis, G\"{o}del did not think that the ``grandfather paradox'' was a compelling reason to discard the possibility of CTCs.  The above quotation continues: ``\emph{\dots  however, in order to prove the impossibility of
    the worlds under consideration, presuppose the actual feasibility of the journey into one’s own past. However, the velocities that would be necessary in order to complete the voyage in a reasonable length of time are far beyond everything that can be
    expected ever to become a practical possibility. Therefore, it cannot be excluded a~priori, on the ground of the argument given, that the space-time structure of the real world [contains CTCs].}''} \qquad --- Kurt G\"{o}del~\cite{God49}
\end{quotation}
In Hawking and Ellis's version, the time-traveler journeys to his past and then prevents himself from embarking on the journey.  
Deutsch modeled this action by a $1$-bit circuit~$C$, consisting entirely of a NOT gate, whose input and output wires were linked by a CTC.
Hawking and Ellis's doubts were based on the G\"{o}delian dilemma that the time-traveling bit canot be~$0$ because $C(0) = 1 \neq 0$, and it cannot be~$1$ because $C(1) = 0 \neq 1$. In other words, there is no $x \in \{0,1\}$ with $C(x) = x$.

But Deutsch's resolution was that there is a \emph{(mixed) quantum state} $\rho = \frac12 \ketbra{0}{0} + \frac12 \ketbra{1}{1}$ with $C(\rho) = \rho$.
Indeed, we don't even ``need'' quantum mechanics to evade the paradox, as long as we allow the universe to have probabilistic states: a consistent possibility for the state of our time-traveling bit~$x$ is ``$x$ equals $0$ or $1$ with probability~$\frac12$ each''.
More generally, if we suppose that a probabilistic (respectively, quantum) circuit~$C$ with~$n$ input and output bits (qubits) implements a Markov chain (quantum channel) of dimension~$2^n$, then there is always an invariant distribution (mixed state) for~$C$, and we get a consistent universe so long as $C$'s input/output are set to this fixed point.

Despite this resolution, Deutsch remained concerned with another philosophical paradox of CTCs, related to the creation of knowledge.  
He imagined a time-traveler~$M$ who takes the proof of a difficult mathematical theorem back in time to the person~$P$ who supposedly first discovered it. 
$P$~receives the proof from~$M$ and publishes it; later, this allows~$M$ to take the published proof back in time to give to~$P$.  
But\dots who thought of the proof?

Deutsch sketched a computational version of this paradox that suggested a CTC capable of sending polynomially many bits a polynomial amount of time into the past could be used to solve $\NP$-complete problems.  
This philosophical anomaly was arguably heightened when Bacon~\cite{Bac04} precisely formalized a ``Deutschian'' model of computation with CTCs and indeed showed it contained the complexity class~$\NP$ (see also work of Brun~\cite{Bru03} and Aaronson~\cite{Aar05}).
Whether this finding should count as evidence against the existence of CTCs is up for debate. %
Nevertheless, in this paper we will mostly take for granted the ``Deutschian'' model of CTC-computation%
\footnote{Except in \Cref{sec:PCTC}, where we will investigate the alternative ``postselecting'' model of CTCs.}
and ask: What computational tasks can it solve when \emph{no} complexity resources (time, space) are imposed?  
That is, what is the \emph{computability theory} of CTCs?

\section{The complexity of probabilistic/quantum programming}
For readers uninterested in time travel, we present an alternative motivation for the problems studied in this paper.
In short, these questions can be viewed as reasoning about the computational complexity of problems associated to infinite-dimensional computable Markov chains --- and their generalization, quantum channels.

Formal reasoning about probabilistic programming has a long history in theoretical computer science.
Already the seminal work of Kozen~\cite{Koz79} recognized the subtlety arising in the analysis of randomized programs that run for an a priori unbounded amount of time, and hence have no finite bound on the number of random variables they involve.
(Indeed, Kozen invoked nontrivial ergodic theory of Banach spaces to give semantics to such programs.)

Formal verification of properties of probabilistic programs has been an active area of research since the early~'80s~\cite{HSP83,SPH84}.
As with deterministic programs, an important challenge is to develop methods for proving termination.
But this is a more nuanced issue when randomness is involved: a probabilistic program might terminate with some probability between $0$~and~$1$; or, it might terminate with probability~$1$ but run for infinitely many steps in expectation.  
See, e.g.~\cite{BG05} for some discussion, including connections with infinite-dimensional Markov chains.

Of course, deciding if a  general deterministic program terminates is the Halting problem, complete for~$\RE$.
Still, one seeks heuristics and restricted cases in which termination is efficiently provable. 
As described in \cite{KKM19}, researchers working on the analogous problem for deciding ``almost sure (probability~$1$) termination'' of probabilistic programs (e.g.~\cite{EGK12,CS13,KKMO16}) found this to be ``more involved to decide''.
And indeed, Kaminski, Katoen, and Matheja~\cite{KKM19} recently showed there is a sense in which this is provably so: they showed that deciding if a given TM~$M$ (running on a blank tape) halts with probability~$1$ is $\Pi_2$-complete; and, deciding if it has finite expected running time is $\Sigma_2$-complete.
Here $\Sigma_2$ is $\RE^{\RE}$, the class of problems semidecidable with a $\Halts$ oracle, and $\Pi_2 = \mathsf{co}\textrm{-}\Sigma_2$.

The results in this work are of a related flavor.
CTCs carrying classical bits give rise to computable Markov chains on a countably infinite state space (and such objects are almost the same thing as TMs running on a blank tape).  
The problem we're concerned with is (approximately) finding an invariant distribution, assuming one exists.
Somewhat surprisingly, we show that this task is complete for $\Delta_2 = \Sigma_2 \cap \Pi_2$; that is, while it's fundamentally harder than the Halting problem, it's \emph{fundamentally easier} than deciding almost sure termination of TMs (and easier, as we also show in \Cref{sec:hardness-classify}, than related Markov chain tasks such as classifying states as positive recurrent/null recurrent/transient).

Our most technical theorem (motivated by qubit-carrying CTCs) extends the above result to computable quantum channels on an infinite-dimensional Hilbert space.  
It shows the same $\Delta_2$-completeness of finding an (approximate) invariant state for a given channel, assuming one exists.
This result requires some very recent quantum Markov map technology (e.g.~\cite{Gir22}), and we suggest formal verification of properties of \emph{quantum} programs as a potentially fruitful avenue for future work.  (See, e.g.~\cite{Ying12} for a start.)

\section{CTC computation definitions and prior work} \label{sec:defs}
The formal definition of the ``Deutschian'' model of computing with closed timelike curves comes from Deutsch~\cite{Deu91}, Bacon~\cite{Bac04}, and Aaronson~\cite{Aar05}. 
Herein we present it directly in terms of Markov chains and quantum channels.

A CTC may be characterized by three properties:
\begin{itemize}
    \item Whether it carries classical bits or qubits. This corresponds to whether the CTC corresponds to a Markov chain or to a quantum channel.
    \item Its \emph{width}~$w$, meaning the number of (qu)bits it can carry.
    The dimensions of the resulting chain/channel are indexed by $W \coloneqq \{0,1\}^w$.\footnote{In fact, there is no special reason to insist that dimensions be a power of~$2$, but we do so for expositional  simplicity; in any case, it is known~\cite{OS14,OS18} not to matter.}
    When $w$ is ``unbounded'', we identify~$W$ with $\{0,1\}^*$.  
    \item Its \emph{length}~$\ell$; i.e., the temporal distance it sends (qu)bits back in time.  
    One way this affects CTC computation is that $O(\ell)$ is an upper bound on the bit-complexity of all entries in the matrix representing the chain/channel.
\end{itemize}
In this work, our chief interest is the case when $w$ and $\ell$ are \emph{unbounded}.

Given a CTC, one can define associated models of computation for deciding whether a string~$x \in \{0,1\}^n$ is in a language (or promise problem)~$L$.
To warm up, in (i)--(iii) below we describe potential models for CTCs carrying finitely many bits.
We'll subsequently discuss the changes needed for CTCs carrying qubits, and then finally our main interest: the case of unboundedly many (qu)bits.
\begin{enumerate}[label=(\roman*)]
    \item On input $x$, an algorithm~$A_1$ gets to output the description of a probabilistic Turing machine\footnote{To be formal, we specify that probabilistic Turing machines ``toss a probability~$1/2$ coin'' on each step.  One could also allow the other standard model, transition probabilities from a fixed finite set of rationals, but in all models studied this won't make any difference; see e.g.\ discussion in~\cite{OS14}.}~$C_x$ that computes a map $\{0,1\}^w \to \{0,1\}^w$.
    This $C_x$ must have the property that it halts with probability~$1$ on all inputs.
    \item This $C_x$ naturally defines a Markov chain on state space~$W = \{0,1\}^w$ with transition matrix~$P_x$. 
    As is well known, there is at least one probability distribution on~$W$ that is invariant for~$P_x$.
    ``Nature'' selects one such invariant distribution~$\pi$ (arbitrarily), and then draws one sample $\bu \sim \pi$.
    \item A second algorithm~$A_2$ takes $x$ and~$\bu$ as input, computes, and makes an accept/reject decision about $x \stackrel{?}{\in} L$.\footnote{
        These steps do not explicitly mention time travel or CTCs (none of our CTC-assisted models will), so let us briefly explain the connection between (i)--(iii) and CTCs at an intuitive level. 
        Suppose a programmer with access to a CTC is trying to solve a computational problem. 
        In (i), they choose the transformations applied to the time-traveling bits (the bits carried by the CTC). 
        In (ii), Nature selects a state $\pi$ that is causally consistent with these transformations---i.e., a fixed point of the CTC. 
        Finally, the programmer receives a sample $\bu\sim\pi$, and, in (iii), performs post-processing to make an accept/reject decision.
        Thus the goal of the CTC programmer is to transform the time-traveling bits so that Nature performs a useful computation on their behalf. 
    }
\end{enumerate}

To fully specify the resulting CTC-assisted complexity class, we need to specify the complexity allowed for algorithms $A_1$, $C_x$, and $A_2$, as well as the allowed probability of error in the final decision about $x \in L$ (zero error, bounded error of~$1/3$, etc.).
Having specified all this, the main question to be answered is 
\[
    \text{``What is the complexity class~$\calF$ of languages decided in the resulting CTC-assisted model?''}
\]

\begin{remark} \label{rem:w}
    Given prior work, the answer to this question about~$\calF$ appears to be primarily controlled by the complexity of~$C_x$ and the value of~$w$ (which, recall, we are so far assuming is finite). The error probability is also sometimes of importance.
    The complexity of $A_1$ and (especially)~$A_2$ are not particularly important, as long as they are not extravagant compared to the complexity of~$C_x$.
    In other words, determining~$\calF$ mainly seems to be about understanding the computational complexity of (approximate) invariant distributions for $W$-state Markov chains computed by programs of $C_x$'s type.
\end{remark}

\subsection{Prior work} \label{sec:prior}
Let us illustrate \Cref{rem:w} while discussing several prior works.

\paragraph{Deterministic $C_x$ with $w = \poly(n)$.} 
Suppose that $w = \poly(n)$, $C_x$ is required to be deterministic $\poly(n)$ time, and zero error is allowed.  
We also assume $A_1$ and $A_2$ are deterministic $\poly(n)$ time. 
In this case, Aaronson and Watrous showed (see~\cite[Sec.~3]{AW09}) the resulting CTC-assisted complexity class~$\calF$ is $\PSPACE$.
For the lower bound $\PSPACE \subseteq \calF$, the basic insight is that the computation of an arbitrary $\PSPACE$ machine~$M$ on an input $x \in \{0,1\}^n$ has only exponentially many possible configurations; hence these configurations are expressible with $w = \poly(n)$ bits. 
Then there is an associated Markov chain on configurations --- which has all transition probabilities equal to $0$ or~$1$ --- that the CTC-assisted computation can build and use to determine the outcome of~$M(x)$. 
Note that in this case, the complexity of~$A_1$ (which uses $x$ to build $C_x$ for the chain) is rather minimal ---  $O(n)$ time in a multitape TM model. 
The complexity of~$A_2$ is even less --- one can arrange for it to be~$O(1)$ time.  
More interestingly, it suffices for the complexity of~$C_x$ to be quite low; \cite{AW09} points out it can be in~$\AC^0$.

The upper bound, $\calF \subseteq \PSPACE$, relies on the following: (a)~that ``deterministic Markov Chains'' have very simple-to-understand invariant distributions, namely the uniform distribution on cycles (and mixtures thereof); (b)~that one can find a vertex in a cycle in an (implicitly represented) $2^{\poly(n)}$-state graph in $\poly(n)$ space; (c)~that the CTC computation's error probability is required to be~$0$. 
Note here that since the upper bound being proven is $\PSPACE$, it is perfectly fine to allow $A_1$, $A_2$, and $C_x$ to run in $\PSPACE$.  
 
\paragraph{Probabilistic $C_x$ with $w = \poly(n)$.} 
Perhaps more natural is to allow bounded error, and~$C_x$ to be a \emph{probabilistic} $\poly(n)$-time algorithm; then one gets $W$-state Markov chains with a wide range of transition probabilities, not just~$0$ and~$1$.
(One might then also allow $A_1$ and~$A_2$ to be probabilistic, but this makes little difference.)
This was indeed the original model considered in~\cite{Aar05} (along with a quantum analogue).  
It turns out that here~$\calF$ is \emph{still} $\PSPACE$, but this new upper bound --- due to Aaronson and Watrous~\cite{AW09} --- is noticeably more sophisticated.  
One needs a $\PSPACE$ algorithm for (approximately) finding an invariant distribution for a Markov chain with an (implicitly specified) transition matrix~$P$ of dimension $W = 2^{\poly(n)}$.  
Aaronson and Watrous give such an algorithm even in the strictly more general case of a quantum channel.
Roughly speaking, they identify a $W$-dimensional operator $\calE$ that projects any initial probability distribution onto its limiting invariant distribution; then they show that $\calE$ can be computed in~$\NC$, hence $\polylog(W) = \poly(n)$ space.  
Incidentally, this relies on the fact that the nonzero entries of~$P$ are at least $\exp(-\poly(n))$, which in turn relies on~$C_x$ being $\poly(n)$-time (which in turn is forced if the CTC has $\poly(n)$ ``length'').
One can also deduce from Aaronson and Watrous's work that $\calC$ is still contained in $\PSPACE$ even if \emph{unbounded} error probability is allowed.%

\paragraph{Probabilistic (and quantum) computation, $w = 1$, and postselection.} As raised in~\cite{AW09}, the case of the narrowest possible CTC, $w = 1$, is particularly interesting. %
As observed by Say and Yakary{\i}lmaz~\cite{SY12}, if the combination of $C_x$'s computation type and the overall error model is complexity class~$\calC$, then the resulting $1$-bit-CTC-assisted class~$\calF$ is generally $\post\calC$, the ``postselected'' version of class~$\calC$.
The notion of ``postselecting'' a class~$\calC$ was first introduced by Aspnes, Fischer, Fischer, Kao, and Kumar~\cite{AFF+01} under the name ``conditional probabilistic computation'' (although it is essentially equivalent to the earlier ``$\mathsf{path}$ operator'' of Hem, Hemaspaandra, and Thierauf~\cite{HHT93}).
It was independently introduced and named ``postselection'' by Aaronson~\cite{Aar04a,Aar05b}.
Known results are $\post\PP = \PP_{\mathsf{path}} = \PP$~\cite{Sim75,AFF+01}, $\post\ZPP = \NP \cap \coNP$~\cite{SY12}, $\post\RP = \NP$~\cite{HHT97}, and $\post\BPP = \BPP_{\path}$ 
(see \cite{HHT93,HHT97,AFF+01,BGM03a,Aar04b}). 
This last class is somewhat lesser known; it satisfies $\MA \cup \PTIME_{\|}^{\NP} \subseteq \BPP_{\path} \subseteq \BPP_{\|}^{\NP}$, and it equals $\PTIME_{\|}^{\NP}$ under a standard derandomization assumption~\cite{SU06}.
As consequences, we get that for $1$-bit CTCs, if $C_x$ is probabilistic $\poly(n)$ time, then the resulting CTC-assisted class~$\calF$ is $\NP \cap \coNP$ (respectively, $\NP$, $\BPP_{\path}$, $\PP$) if the model has zero (respectively, bounded $1$-sided, bounded $2$-sided, unbounded\footnote{Bounded $1$-sided error: ``yes'' inputs accepted with probability exceeding~$2/3$, no inputs accepted with probability equal to~$0$.  Bounded $2$-sided error: change ``equal to~$0$'' to ``less than~$1/3$''.  Unbounded $2$-sided error: change both $2/3$ and $1/3$ to $1/2$.}) error.
Finally, Aaronson~\cite{Aar04a} proved that $\post\BQP = \PP$.  
This implies~\cite{SY12} that if we have a (still \emph{classical}) $1$-bit CTC, the four Markov chain transition probabilities are computed by a $\poly(n)$-time \emph{quantum} algorithm~$C_x$\footnote{\label{foot1}That is, $C_x$ is a unitary circuit defined by $A_1(x)$, it takes in one qubit in a computational basis state, it uses $\poly(n)$ ancillas and gates from a fixed rational universal gateset, and its classical output is given by measuring its first wire and discarding the remaining qubits.}, and we have bounded $2$-sided error, then the resulting CTC-assisted class~$\calF$ is $\PP$.

\paragraph{Logarithmic $w$.}  One more natural setting is $w = O(\log n)$, meaning the CTC's Markov chain has $\poly(n)$ states.  
It was shown in \cite{OS14} that all the resulting classes~$\calF$ are unchanged from their $w = 1$ counterparts.

\subsection{Qubit-carrying CTCs}

We now discuss how to modify definitions to accommodate CTCs carry qubits (as in the original model of Deutsch~\cite{Deu91}).
Here, the $W$-state Markov chain $P_x$ gets replaced by a \emph{quantum channel} (completely positive, trace-preserving linear map) $\Psi_x$, mapping mixed states on~$\C^{W}$ to mixed states on~$\C^W$.
Continuing to assume $W$ is finite, any such channel $\Psi_x$ has at least one invariant state\footnote{In quantum contexts, ``state'' will henceforth always mean ``mixed state''.}, and the model of CTC-assisted computation is as in (i)--(iii) above, except that rather than $\bu$, algorithm $A_2$ gets one copy of an invariant state~$\rho$ for~$\Psi_x$.

Clearly, the algorithms $C_x$ and~$A_2$ must now be capable of operating on qubits.
(As for algorithm~$A_1$, it is more like the ``uniformity algorithm'' for~$C_x$, and it's typically sufficient/natural for it to be deterministic $\poly(n)$ time.) 
It is perhaps most natural to allow~$C_x$ and $A_2$ to be $\poly(n)$-sized ``general'' quantum circuits~\cite{Wat11}, meaning unitary circuits (as in \Cref{foot1}) with ancilla and measurement gates.

Somewhat interestingly, although quantum channels strictly generalize Markov chains, we are not aware of any natural CTC-assisted model in which the ability to send qubits instead of bits fundamentally increases computational power. 
For instance, as mentioned, Aaronson and Watrous~\cite{AW09} showed that when $w = \poly(n)$, the resulting complexity class~$\calF$ is still $\PSPACE$. 
Similarly, it was shown in~\cite{OS18} that CTCs with qubit width $w = O(\log n)$ --- i.e., quantum channels of $\poly(n)$ dimension --- still just allow for $\calF = \PP$.

\subsection{Infinite-dimensional chains and channels}   \label{sec:infinite}

Now we come to the subject of the present paper: CTCs of unbounded width and length.
Again, let us first focus on the case of CTCs carrying classical bits; i.e., Markov chains of (countably) infinite dimension.

A significant difference between the finite- and infinite-dimensional cases is that a Markov chain on an infinite set of states need not have \emph{any} invariant distribution.
For a simple family of examples, fix $p \in [0,1]$ and consider the Markov chain with state space~$\Z$ in which $u \in \Z$ transitions to $u+1$ with probability~$p$ and to $u-1$ with probability~$1-p$.
It is easy to show that there is no probability distribution on~$\Z$ left invariant by this chain (although arguably this is for two different ``reasons'' depending on whether or not $p = 1/2$ and hence whether the chain is transient or null recurrent; see \Cref{sec:markov-prelim}).

In light of this, to define CTC-assisted computation, we simply stipulate that the Markov chain induced by the CTC have at least one invariant distribution. %
Thus the model of computation will be fixed as follows:
\begin{enumerate}[label=(\roman*)]
    \item On input $x$, a deterministic algorithm~$A_1$ must halt and output the description of a probabilistic Turing machine~$C_x$ that computes a map $\{0,1\}^* \to \{0,1\}^*$.
    This $C_x$ must have the property that it halts with probability~$1$ on all inputs.
    Moreover, the Markov chain~$P_x$ on $\{0,1\}^*$ induced by~$C_x$ must have at least one invariant distribution.
    \item ``Nature'' selects an invariant distribution~$\pi$ for $P_x$ (arbitrarily), and then draws one sample $\bu \sim \pi$.
    \item A second probabilistic TM~$A_2$ (required to halt with probability~$1$ on all inputs) takes $x$ and~$\bu$ as input, computes, and makes an accept/reject decision about $x \stackrel{?}{\in} L$.
\end{enumerate}

Except for the issue of error probability, this completely defines the model of classical CTC-assisted computation~$\calF$ we will study in this work.
Our first main theorem will be:
\begin{theorem} \label{thm:main-classical}
    In our model of classical CTCs of unbounded width and length, regardless of whether $\calF$ is defined with zero error, bounded error, or unbounded error, we have $\calF = \Delta_2$.%
\end{theorem}

Next we discuss computation with qubit-carrying CTCs of unbounded width and length. 
These induce quantum channels over Hilbert spaces of (countably) infinite dimension.
As these strictly generalize infinite-dimensional Markov chains, we again have the issue that there need not be an invariant state; so, we again impose the requirement that the channel have at least one invariant state.
Thus our model is:
\begin{enumerate}[label=(\roman*)]
    \item On input $x$, a deterministic algorithm~$A_1$ must halt and output the description of a general quantum Turing machine~$C_x$ (with the ability to add ancillas and do measurements) that computes a quantum channel from $\ell_2(\{0,1\}^*)$ to $\ell_2(\{0,1\}^*)$.
    This $C_x$ must have the property that it halts with probability~$1$ on all inputs.
    Moreover, the quantum channel~$\Psi_x$ induced by~$C_x$ must have at least one invariant %
    state.
    \item ``Nature'' selects an invariant state~$\rho$ for $\Psi_x$ (arbitrarily).
    \item A second general quantum TM~$A_2$ (required to halt with probability~$1$ on all inputs) takes $x$ and~$\rho$ as input, computes, and makes an accept/reject decision about $x \stackrel{?}{\in} L$.
\end{enumerate}
Before addressing the question of precisely defining the quantum Turing machine model, we state our second main theorem:
\begin{theorem} \label{thm:main-quantum}
    In our model of quantum CTCs of unbounded width and length, regardless of whether $\calF$ is defined with zero error, bounded error, or unbounded error, we have $\calF = \Delta_2$.
\end{theorem}
In particular, once again we find that allowing CTCs to carry qubits rather than bits does not allow for more computational power.

Returning to the definition of quantum Turing machines:  Even the most thorough definition of quantum TMs we know of (perhaps~\cite{Wat99})
does not address quantum TMs that: (a)~may run for an unbounded amount of time; (b)~may take non-unitary actions.
Thus it might seem our definition is a little underspecified.  
To avoid a lengthy digression into quantum TM definitions, we will continue to leave it underspecified, subject to the following explanations.
First, the lower bound $\Delta_2 \subseteq \calF$ in \Cref{thm:main-quantum} is already covered by the lower bound in \Cref{thm:main-classical}, since any reasonable definition of computable quantum channels will include computable Markov chains as a special case. 
Second, for the upper bound $\calF \subseteq \Delta_2$ in \Cref{thm:main-quantum}, all we will need is the following basic assumption, which will be satisfied by any reasonble definition of computable quantum channels:

\paragraph{Assumption.} \emph{A computable quantum channel over $\ell_2(\{0,1\}^*)$ has the following property: There is a deterministic TM that, given as input the classical description of a finitely supported state $\rho$ on $\ell_2(\{0,1\}^*)$ having entries from $\Q[\mathrm{i}]$, as well as a rational $\eps > 0$, outputs the classical description of a finitely supported matrix~$\sigma$ with entries from $\Q[\mathrm{i}]$ satisfying $\trnorm{\sigma - \Psi(\rho)} \leq \eps$.}

\begin{remark}
    As followup to this work, Raat~\cite{Raa23} showed that in a model where CTCs are allowed to \emph{nest} up to depth~$d$, \Cref{thm:main-classical,thm:main-quantum} have extensions in which~$\Delta_2$ is replaced by~$\Delta_{d+1}$.
\end{remark}

\section{Preliminaries on Markov chains, quantum channels, and the arithmetical hierarchy}

\subsection{Markov chains}  \label{sec:markov-prelim}
We briefly recap some basic theory of Markov chains with countably infinite state spaces (see, e.g., \cite{Nor98,Por24}).

We take the set of states to be~$\N$, without loss of generality.  
For a vector $v \in \R^\N$ we write $\trnorm{v} = \sum_{j \in \N} \abs{v_j}$.
A probability distribution on states~$\pi$ is considered to be a nonnegative \emph{row} vector in~$\R^\N$ with~$\trnorm{\pi} = 1$.
A Markov chain is defined by its transition operator~$P \in \R^{N \times N}$; this is any stochastic matrix, meaning one with nonnegative entries and rows summing to~$1$.  
We remark that~$P$ is contractive with respect to~$\trnorm{{\cdot}}$: $\trnorm{vP} \leq \trnorm{v}$ always.
An \emph{invariant distribution} for the Markov chain is a distribution~$\pi$ satisfying $\pi P = P$.
In general, $P$ need not have any invariant distribution.

For $x \in \N$, let $T_x$ denote the $(\N^+ \cup \{\infty\})$-valued random variable counting the number of steps it takes for the Markov chain to first return to~$x$ when starting from~$x$.
Now the set of states for the Markov chain is partitioned as $\N = T \sqcup R_0 \sqcup \R_+$, where:
\begin{itemize}
    \item $T$ is the set of \emph{transient} states, the $x \in \N$ for which $\Pr[T_x < \infty] < 1$.
    \item $R_0$ is the set of \emph{null recurrent} states, the~$x \in \N$  which have $\Pr[T_x < \infty] = 1$ but $\E[T_x] = \infty$.
    \item $R_+$ is the set of \emph{positive recurrent} states, the $x \in \N$ which have $\E[T_x] < \infty$.
\end{itemize}
All invariant distributions for a chain are supported on its positive recurrent states; indeed, an alternative definition for a state~$x$ being positive recurrent is that there exists an invariant distribution with nonzero probability mass on~$x$.

By a computable Markov chain, we refer to a Markov chain induced by a probabilistic Turing machine that halts with probability~$1$ on all inputs.

\subsection{Quantum channels}
Let $\calH$ be a separable Hilbert space.
A (mixed, normal) quantum state~$\rho$ is a positive semidefinite operator on~$\ell_2$ with $\trnorm{\rho} = 1$, where $\trnorm{{\cdot}}$ is the trace norm defined by $\trnorm{X} = \tr\sqrt{X^\dagger X}$.
We write $\calB_1(\calH)$ for the trace class operators on~$\calH$, meaning those linear operators $X : \calH \to \calH$ with $\trnorm{X} < \infty$.
A quantum channel on~$\ell_2$ will mean a linear operator~$\Psi$ on $\calB_1(\calH)$ that is completely positive and trace-preserving. 
Equivalently, $\Psi(X) = \sum_j M_j X M_j^\dagger$ for some sequence of $M_j \in \calB(\calH)$ with $\sum_j M_j^\dagger M_j = \Id$, where $\calB(\calH)$ denotes the set of bounded linear operators on $\calH$.
Such channels are again contractive with respect to~$\trnorm{{\cdot}}$: $\trnorm{\Psi(X)} \leq \trnorm{X}$ always.
As with Markov chains, we often think of taking a starting state~$\rho_0$ and repeatedly applying the channel, forming $\rho_t = \Psi^t(\rho_0)$.
An \emph{invariant state} for~$\Psi$ is a quantum state~$\rho$ with $\Psi(\rho) = \rho$; but again, $\Psi$ need not have an invariant state in general.

A theory of transient, null recurrent, and positive recurrent subspaces for an infinite-dimensional quantum channel~$\Psi$ has developed over the last couple of decades; see, e.g.,~\cite{FR03,Fag04,Uma06,GK12,BN12,CP16,CP16b,CG21,Gir22}. 
This is an orthogonal decomposition $\calH = \calT \oplus \calR_0 \oplus \calR_+$ induced by~$\Psi$, which we briefly define.
First, the positive recurrent subspace $\calR_+$ may be defined as all those~$\ket{x}$ that are in the support of some invariant state for~$\Psi$.
It remains to define the transient subspace~$\calT$, as then the null recurrent subspace~$\calR_0$ may be defined as the orthogonal complement of $\calT \oplus \calR_+$.

Recall that a Markov chain state~$x$ is transient if the expected number of visits to~$x$ when starting from~$v$ is bounded uniformly in~$v$. 
The definition of the transient subspace for a quantum channel generalizes this.
Consider for a positive $A \in \calB(\calH)$ the quadratic form on~$\calH$ defined by
\begin{equation}
    \mathfrak{U}(A)[\ket{v}] \coloneqq \sum_{t = 0}^\infty \tr(\Psi^t(\ketbra{v}{v}) A) \in [0,\infty].
\end{equation}
(If~$A$ is the projector onto some subspace~$X$, one may think of this as the ``expected number of visits to~$X$ when starting from~$\ket{v}$''.)
If the set of~$\ket{v}$ for which $\mathfrak{U}(A)[\ket{v}] < \infty$ is dense, then $\mathfrak{U}(A)$ can be represented by a bounded PSD operator $\mathcal{U}(A)$.  
Then the transient subspace~$\calT$ can be defined to be all those~$\ket{x}$ in the support of some such $\mathcal{U}(A)$.

\subsection{The arithmetical hierarchy}

We will characterize the computational power of the CTC-assisted model with unbounded width and length using classes from the arithmetical hierarchy. For a detailed overview of the arithmetical hierarchy, see \cite{Soa87}. Here we only briefly describe the classes of the hierarchy that we use in this paper.

The arithmetical hierarchy is the union of a sequence of classes $\Sigma_0 \subsetneq \Sigma_1 \subsetneq \Sigma_2 \subsetneq \cdots$.
These are defined as follows: $\Sigma_0 = \Rec$, the decidable languages; and, $\Sigma_{i+1}$ is the set of languages accepted (i.e., semi-decided) by a Turing Machine with an oracle for~$\Sigma_i$.
One also defines $\Pi_i$ as the complement of $\Sigma_i$ and $\Delta_i = \Sigma_i \cap \Pi_i$; the latter is equivalent to the languages decidable by a TM with an oracle for $\Sigma_{i-1}$.  
We will be particularly interested in~$\Delta_2 = \Rec^{\Halts}$.
In \Cref{sec:PCTC} we will also briefly make reference to a class from Ershov's hierarchy~\cite{Ers68}, namely the ``$\omega$-c.e.\ languages''.  
We will notate this class as $\Rec_{\|}^{\Halts}$, since it is also known~\cite{Car72} to be equivalent to the class of languages that truth-table reduce to the Halting problem.
Ershov showed that $\Rec_{\|}^{\Halts} \subsetneq \Delta_2$.

\section{Our lower bound for CTC-assisted computation}   \label{sec:class-lower}
In this section we prove the lower bound in \Cref{thm:main-classical}, namely:
\begin{theorem} \label{thm:main-lower}
    Every language $L \in \Delta_2$ is decidable with zero error in the classical CTC-assisted computation model of \Cref{sec:infinite}.
\end{theorem}
As a warmup, we first prove:
\begin{proposition} \label{prop:halts}
    \Cref{thm:main-lower} holds for $L = \Halts$. 
\end{proposition}
\begin{proof}
    Let $P_\bot$ denote the following Markov chain on~$\N$:  each state $u \in \N$ transitions to $u+1$ with probability~$1/2$ and to~$0$ with probability~$1/2$.
    It is easy to show $P_\bot$ has an invariant distribution (indeed, it has a unique one, $(1/2,1/4,1/8, \dots)$).
    Let $P^{(u)}_\bot$ be the variant of $P_\bot$ in which the transitions out of state~$u$ are replaced with a self-loop. 
    It is easy to show that $P^{(u)}_\bot$ has as its unique invariant distribution as the one that puts 100\% probability on~$u$.

    Now to decide whether an input Turing Machine~$X$ halts (on blank input tape), our algorithm~$A_1$ will prepare a Turing machine~$C_X$ that implements $P^{(v)}_\bot$ if $X$ halts precisely on the $v$th time step, and implements $P_\bot$ otherwise.
    (There is indeed an easy $C_X$~that implements this: on input~$u$, it simply simulates~$X$ for~$u$ steps to see if it halts precisely at time~$u$.  
    If so, it outputs~$u$; otherwise it outputs $0$ or $u+1$ with probability~$1/2$ each.)
    Finally, upon receiving sample~$\bv$ from the invariant distribution of the chain defined by $C_X$, our algorithm~$A_2$ simulates~$X$ for $\bv$ steps and accepts if and only if~$X$ has halted by then.  

    Obviously, if $X$ does not halt then the overall algorithm can never wrongly accept.  
    On the other hand, if~$X$ halts at some time~$v$, then $C_X$ will implement $P^{(v)}_\bot$, so~$\bv$ will equal~$v$ with certainty and~$A$ will always accept.
\end{proof}

We now prove \Cref{thm:main-lower}.
\begin{proof}
    Let $M_1, M_2, \dots$ be a computable enumeration of all Turing machines, and let $h_j \in \N \cup \{\infty\}$ denote first time step on which $M_j$ halts (when run on a blank tape), or $\infty$ if $M_j$ does not halt.  
    Also, define $f : (\N \cup \{\infty\}) \to \{0,1\}$ by $f(H) = 1$ iff $H \neq \infty$, so that $f(h_j)$ is the $0$-$1$ indicator for whether $M_j$ halts.

    Let us say that a TM is augmented with a ``special'' tape if it gets access to a read-only tape in which, at the $j$th time step, the bit $f(h_j)$ is appended to the end of the tape.
    It is an exercise to show that every language $L \in \Delta_2$ reduces to the task of deciding --- given a TM $X$ with special tape, promised to halt on a blank input --- whether~$X$ accepts or rejects.%
    \footnote{This exercise is similar to the well-known fact that computing Chaitin's constant~$\Omega$~\cite{Cha75} is $\Delta_2$-hard (indeed, it's $\Delta_2$-complete).%
    }
    We now describe our CTC-assisted way of solving this problem.

    Given $X$, the algorithm~$A_1$ will construct a TM $C_X$ implementing a certain Markov chain.
    It will be clear from our description of~$X$ that an appropriate algorithm~$A_1$ exists. 
    The chain's state set will be (an appropriate encoding of) $S_0 \sqcup S_1 \sqcup S_2 \sqcup \cdots$, where $S_k$ is a copy of the set $\N^k$.
    We refer to the states in $S_k$ as ``$k$-tuples''.
    
    The chain restricted to $S_k$ will somewhat resemble a $k$-fold product of the chains used in the proof of \Cref{prop:halts}.
    Specifically, on input state $u = (u_1, u_2, \dots, u_k) \in S_k$, the transition algorithm $C_X$ will do the following:
    \begin{enumerate}
        \item For each $j \in [k]$, run $M_j$ to see if it halts precisely on the $u_j$th step; i.e., if $h_j = u_j$.
        If so, define bit $g_j = 1$; otherwise, define $g_j = 0$.
        (This is a ``guess'' for the true value $f(h_j)$.)
        \item Simulate $X$ for $k$ steps, using $g_j$ as the $j$th value placed on its special tape.
        \item If $X$ has not halted by then, $C_X$ will output $(0, 0, \dots, 0) \in S_{k+1}$.
        \item Otherwise, $C_X$ acts component-wise on $u$ as in \Cref{prop:halts}.
        That is, it outputs $(\bu'_1, \dots, \bu'_k) \in S_k$, where:
        \begin{equation}
            \bu'_j = \begin{cases}
                        u_j, & \text{if $g_j = 1$;}\\
                        0 \text{ or } u_j+1 \text{ with probability $1/2$ each,} & \text{if $g_j = 0$.}
                     \end{cases}
        \end{equation}
    \end{enumerate}
    Finally, upon receiving some sample~$\bv = (\bv_1, \bv_2, \dots, \bv_{\bell})$ from an invariant distribution for the chain, algorithm $A_2$ performs the same Steps 1--2 above as $C_X$, and then accepts or rejects as $X$ does.
    (We will argue that $X$ \emph{always} halts within~$\bell$ steps in $A_2$'s simulation.)

    We will now argue two things to complete the proof.
    First, we'll show that the chain defined by~$C_X$ always has at least one invariant distribution, as required.
    Second, we'll show that any tuple $v = (v_1, \dots, v_m)$ in the support of an invariant distribution must be a \emph{valid tuple}, meaning one with the following two properties:
    \begin{enumerate}[label=(\roman*)]
        \item $v_j = h_j$ for all $j \in [m]$ with $h_j \neq \infty$;
        \item $m$ is at least the true running time of~$X$.
    \end{enumerate}        
    This will indeed complete the proof, as it is clear that $A_2$ will simulate~$X$ successfully whenever it is given a valid tuple (in particular, its $g_1, \dots, g_m$ bits will equal $f(h_1), \dots, f(h_m)$).

    Let us first show that the chain indeed has an invariant distribution. 
    Let $m$ be an upper bound on the running time of~$X$. 
    We exhibit an invariant distribution $\pi$ supported on~$S_m$.  
    The distribution will simply be the product distribution $\pi = \pi_1 \times \pi_2 \times \cdots \times \pi_m$, where $\pi_j$ is 100\% concentrated on~$h_j$ when $h_j \neq \infty$, and otherwise $\pi_j$ is the unique invariant distribution for $P_\bot$ (from \Cref{prop:halts}) when $h_j = \infty$.
    To see that this distribution~$\pi$ is invariant, we observe how $C_X$ acts under it. 
    First, by construction it is easy to see that any $\bv \sim \pi$ will be a valid tuple.
    This means that~$C_X$ will never output $(0, 0, \dots, 0) \in S_{m+1}$ in line~3 of its definition; rather, it will always apply line~4.
    Moreover, the validity implies that $C_X$ acts just as in \Cref{prop:halts} for each of the $m$~components in the tuple, confirming that~$\pi$ is indeed invariant.

    It remains to verify that \emph{every} invariant distribution for the chain has all its support on valid tuples. Since every invariant distribution is supported on the positive recurrent states of the chain (as explained in \Cref{sec:markov-prelim}),
    it suffices to show that every \emph{invalid} tuple is not positive recurrent.
    In fact, we show invalid tuples are transient.

    To show this, we need to show that if the chain is started from an invalid tuple~$w = (w_1, \dots, w_k)$, the expected total number of returns to~$w$ (over the whole trajectory) is finite.
    (In fact, in many cases we will show that there are \emph{no} returns to~$w$.)
    Let us consider two cases for why~$w$ is invalid. 
    First, suppose condition~(i) holds but condition~(ii) fails.
    In this case, when $C_X$ gets input~$w$, all its $g_j$ bits will match the correct values $f(h_j)$, and hence $C_X$ will accurately simulate~$X$ for~$k$ steps.  But since (ii)~fails, this is not enough time for~$X$ to halt, so $C_X$ will output $(0, 0, \dots, 0) \in S_{k+1}$.
    Now observe that once the chain reaches~$S_{k+1}$, it can never return to~$S_k$. 
    Thus $w$ is certainly a transitive state in this case.

    It remains to check that $w$ is transient in the case that $w$ is invalid because condition~(i) fails.
    In this case, there is at least one $J \in [k]$ with $h_J \neq \infty$ and $w_J \neq h_J$.
    Now in considering whether there is a finite number of expected returns to~$w$, we may assume that line~3 is deleted from algorithm~$C_X$, because if ever the chain enters~$S_{k+1}$, it will have no more visits to~$w$.
    But with this deletion, the chain behaves exactly as
    $P_{\bot}^{(h_J)}$ (from \Cref{prop:halts}) when restricted to just its $J$th component.
    Since $w_J$ is easily seen to be a transient state for $P_{\bot}^{(h_J)}$ (recall $w_J \neq h_J$, the unique recurrent state in $P_{\bot}^{(h_J)}$), 
    we conclude that there are indeed only finitely many returns to~$w$ in expectation, since there are only finitely many times when the state's $J$th component equals~$w_J$.
    This completes the proof that all invalid~$w$ are transient.
\end{proof}

\section{Our upper bound for classical CTC-assisted computation} \label{sec:class-upper}
In this section we prove the upper bound in \Cref{thm:main-classical}, namely:
\begin{theorem} \label{thm:classical-upper}
    Every language decidable with unbounded error in the classical CTC-assisted model from \Cref{sec:infinite} is in~$\Delta_2$.
\end{theorem}
Given that we will later prove the strictly stronger \Cref{thm:main-quantum}, upper bounding the power of quantum CTCs, it is formally redundant to prove \Cref{thm:classical-upper}.
However we believe that proving it first helps clarifies the ideas underlying our work.

\bigskip
The essential result underlying \Cref{thm:classical-upper} is the following one about deciding whether a distribution is close to an invariant distribution for a given Markov chain:
\begin{theorem} \label{thm:class-approx}
    Given a computable Markov chain on~$\N \cong \{0,1\}^*$ and a distribution over $\N$, the task of deciding whether the distribution is close to an invariant distribution for the chain many-one reduces to $\cHalt$. %
    More precisely, there is an algorithm~$F$ with the following properties:
    \begin{itemize}
        \item $F$ takes three inputs: a Turing machine~$C$ implementing a Markov chain~$P$ on state space $\N$; a finitely supported rational probability vector~$\wh{\pi}$ on~$\N$; and, a rational $\eps > 0$.
        \item $F$ outputs a TM~$Y$.
        \item If $\trnorm{\wh{\pi} - \pi} \leq \eps$ for some invariant distribution~$\pi$ for~$P$, then $Y \in \cHalt$.
        \item If $Y \in \cHalt$, then there is an invariant distribution~$\pi$ for~$P$ satisfying $\trnorm{\wh{\pi} - \pi} \leq \eps' = 6\eps$.
    \end{itemize}
\end{theorem}

\begin{remark}
    The precise function $\eps'$ of~$\eps$ here is not important here; all we need is that $\eps' \to 0$ as $\eps \to 0$.
\end{remark}

\subsection{Deducing \Cref{thm:classical-upper} from \Cref{thm:class-approx}}
Before we begin, a simple lemma:
\begin{lemma}   \label{lem:comp}
    There is an algorithm~$B$ with the following property: Given as input a TM $C$ implementing Markov chain $P$ on~$\N$, a rational probability vector $\pi$ of finite support, and a rational $\eps > 0$, the algorithm~$B$ outputs a rational probability vector $\sigma$ of finite support satisfying $\norm{\sigma - \pi P}_1 \leq \eps$.
\end{lemma}
\begin{proof}
    We may assume $\eps \leq 2$, else the problem is trivial.

    We first consider the task of approximately computing one row of~$P$, say $P_{u{\cdot}}$.
    Assuming without loss of generality that $C$ flips a (fair) coin at every time step, there is a deterministic algorithm that, on input~$u \in \N$, computes (level-by-level) the full binary tree corresponding to $C(u)$'s computation (in which each node at level~$t$ would occur with probability~$2^{-t}$).
    
    Since $C(u)$ halts with probability~$1$, there must be some time/level~$t_u$ such that $C(u)$ halts with probability at least $1-\eps/4$ by time~$t_u$.  
    It follows there is a deterministic, halting TM~$B_1$ that, on input $u \in \N$ and~$\eps$, outputs a nonnegative dyadic vector of finite support, $p'_u \leq P_{u \cdot}$, with $\norm{p'_u - P_{u \cdot}}_1 \leq \frac{\eps}{4}$.

    With $B_1$ in hand, we can describe~$B$.  
    On input $\pi$ of finite support~$U \subset \N$, algorithm~$B$ can compute~$p'_u$ for all $u \in U$ (using~$B_1$), and then compute $\wh{\sigma} = \sum_{u \in U} \pi_u p'_u$, a nonnegative rational vector of finite support.
    Since $\pi P = \sum_{u \in U} \pi_u P_{u \cdot}$, one easily concludes $\norm{\wh{\sigma} - \pi P}_1 \leq \frac{\eps}{4}$.
    It remains to adjust $\wh{\sigma}$ to a probability vector, which causes little error since we now know $\norm{\wh{\sigma}}_1 \geq 1-\frac{\eps}{4}$.
    Specifically, $B$~will finally output $\sigma = \wh{\sigma}/\norm{\wh{\sigma}}_1$, which is easily checked to satisfy $\norm{\sigma - \wh{\sigma}}_1 \leq \frac{3\eps}{4}$, completing the proof.
\end{proof}

With this lemma in hand, we explain why \Cref{thm:class-approx} implies \Cref{thm:classical-upper}.
\begin{proof}[Proof of \Cref{thm:classical-upper} from \Cref{thm:class-approx}.]
Suppose $L$ is decided by the CTC-assisted algorithms $A_1$ and~$A_2$ as in \Cref{sec:infinite}.
We describe an algorithm with a $\cHalt$ oracle (i.e., a $\Delta_2$-algorithm)~$M$ that decides~$L$.
Of course, this $M$ will often use the algorithm~$F$ from \Cref{thm:class-approx} on a given distribution~$\wh{\pi}$ and tolerance parameter~$\eps$, and then apply its $\cHalt$ oracle to the result.  
We will refer to this as ``\emph{Deciding if $\wh{\pi}$ is $(\eps,\eps')$-close to invariant}''.

On input~$x$, algorithm $M$ first uses $A_1$ to create the probabilistic TM $C = C_x$ implementing Markov chain $P = P_x$ with at least one invariant distribution.
What we know is that for any invariant distribution~$\pi$ for~$P$, the  number
\begin{equation}    \label{eqn:L}
    p(x,\pi) \coloneqq \Pr_{\bv \sim \pi}[A_2(x,\bv) \text{ accepts}] \neq 1/2
\end{equation}
is greater than $1/2$ iff $x \in L$.
As the action of $A_2(x, \cdot)$ is, formally, a computable Markov chain on~$\N$ (where all transitions go to $0 = \text{reject}$ or $1 = \text{accept}$), it follows from \Cref{lem:comp} that for any finitely supported rational distribution~$\pi'$ and any rational tolerance $\eps_a > 0$, algorithm~$M$ can approximate $p(x,\pi')$ to within an additive~$\pm \eps_a$.

\medskip

Algorithm~$M$ will now act as follows:

\medskip

\qquad In a dovetailing fashion, loop over all rational probability vectors $\wh{\pi}$ on~$\N$ and rational $0 < \eps_a, \eps_c < 1$:

\qquad \qquad Decide if $\wh{\pi}$ is $(\eps_c,\eps'_c)$-close to invariant. If not, continue to the next $(\wh{\pi}, \eps_a, \eps_c)$.

\qquad \qquad Else if so, compute an estimate $\wh{p}(x,\wh{\pi})$ that is within $\pm \eps_a$ of $p(x,\wh{\pi})$, and then\dots

\qquad \qquad \qquad $M$ accepts if $\wh{p}(x,\wh{\pi}) > 1/2 + \eps_a + \eps'_c$; rejects if $\wh{p}(x,\wh{\pi}) < 1/2 - \eps_a - \eps'_c$; else continues.
    
\medskip

We need to show that $M$ always terminates and is always correct.  
Starting with correctness, suppose $M$ terminates after discovering some $\wh{\pi}$  such that $\abs{\wh{p}(x,\wh{\pi}) - 1/2} > \eps_a + \eps'_c$.
First, we have $\abs{p(x,\wh{\pi}) - 1/2} > \eps'_c$.
Second, since the $\cHalt$ oracle accepted, the guarantee from \Cref{thm:class-approx} is that there is an invariant distribution~$\pi$ for~$P$ with $\trnorm{\wh{\pi} - \pi} \leq \eps'_c$.
Thus $\abs{p(x,\wh{\pi}) - p(x,\pi)} \leq \frac12 \eps'_c < \eps'_c$.
We conclude that $p(x,\wh{\pi})$ and $p(x,\pi)$ are on the same side of~$1/2$, and hence $M$'s accept/reject decision is correct.

As for showing termination on input~$x$, fix some invariant distribution $\pi$ for~$P$. 
Since $p(x,\pi) \neq 1/2$, we can select $\eps > 0$ such that $\abs{p(x,\pi) - 1/2} > 2\eps + \eps'$ (recalling $\eps' = 3\eps \to 0$ as $\eps \to 0$).
Then let $\wh{\pi}$ be a rational probability vector of finite support satisfying $\trnorm{\wh{\pi} - \pi} \leq \eps$.
Now it is easy to see that $M$ will halt whenever it reaches $\wh{\pi}$ and $\eps_a, \eps_c < \eps$.
\end{proof}

\subsection{Reducing \Cref{thm:class-approx} to a theorem on everlasting-invariance}
Given a computable Markov chain~$P$ with at least one invariant distribution, a normal algorithm can only easily find ``approximately invariant'' distributions, meaning $\wh{\pi}$ satisfying $\trnorm{\wh{\pi} - \wh{\pi}P} \leq \eps$.
However, in infinite-dimensional Markov chains, approximately invariant distributions can be arbitrarily far away from genuinely invariant distributions.
That said, let us consider the following notion:
\begin{definition}
    If $P$ is the transition matrix of a Markov chain, we say distribution $\wh{\pi}$ is \emph{$\eps$-everlasting-invariant} if $\trnorm{\wh{\pi} - \wh{\pi}P^t} \leq \eps$ for all $t \in \N$.
\end{definition}
The two key insights underlying our proof of \Cref{thm:class-approx} are: 
\begin{enumerate}
    \item Deciding if a distribution is $\eps$-everlasting-invariant reduces to $\cHalt$.
    \item Any $\eps$-everlasting-invariant distribution \emph{is} close to a truly invariant distribution.
\end{enumerate}
Insight~1 above is straightforward, given the definition of everlasting-invariance.  
Insight~2 is more technically interesting, and requires a modestly good understanding of the theory of Markov chains on infinitely many states (a topic which, luckily, has been well-understood for 50+ years).

We start with the easier observations.
First, because Markov chain transition matrices are contractive in $1$-norm, we have the following:
\begin{fact}    \label{fact:class-easy}
    If distribution $\wh{\pi}$ satisfies $\trnorm{\wh{\pi} - \pi} \leq \eps$ for some invariant distribution~$\pi$ for~$P$, then $\wh{\pi}$ is $\eps$-everlasting-invariant.
\end{fact}

Next, we verify Insight~1 above:
\begin{proposition} \label{prop:class-approx}
    A variant of \Cref{thm:class-approx} holds, with the following two conclusions:
    \begin{itemize}
        \item If $\wh{\pi}$ is $\eps$-everlasting-invariant, then $Y \in \cHalt$.
        \item If $Y \in \cHalt$, then $\wh{\pi}$ is $2\eps$-everlasting-invariant.
    \end{itemize}
\end{proposition}
\begin{proof}
    Observe that our algorithm~$F$ for this task is able to produce the description of TMs $B_1, B_2, \dots$, where $B_t$ uses \Cref{lem:comp} on $C^t$ (the TM for~$P^t$) to compute a rational probability vector $\sigma_t$ satisfying $\trnorm{\sigma_t - \wh{\pi}P^t} \leq \frac{\eps}{2}$.
    Now~$F$ simply outputs the TM~$Y$ which loops over all $t \in \N$, runs~$B_t$, and halts if it ever discovers that $\trnorm{\sigma_t - \wh{\pi}} > \frac{3\eps}{2}$.
    It is easy to check using the triangle inequality that the two required conclusions hold.
\end{proof}

In light of \Cref{prop:class-approx} and \Cref{fact:class-easy}, we see that to prove \Cref{thm:class-approx} (and hence our upper bound of~$\Delta_2$ for classical CTC computation, \Cref{thm:classical-upper}), it remains to show the following:  If $\wh{\pi}$ is $2\eps$-everlasting-invariant, then there is some truly invariant distribution within~$6\eps$ of~$\wh{\pi}$.
This is a purely technical result about Markov chains, and we establish it in the next section.

\subsection{Our main Markov chain result} \label{sec:final}
As described above, to complete the proof of \Cref{thm:classical-upper}, it suffices for us to show the following theorem on Markov chains:
\begin{theorem} \label{thm:mc}
    Let $P$ be the transition operator for a Markov chain on~$\N$.
    Suppose that $\wh{\pi}$ is $\eps$-everlasting-invariant for~$P$, where $\eps < 1$.
    Then there is an invariant distribution~$\pi$ for~$P$ with $\norm{\wh{\pi}- \pi}_1 \leq 3\eps$.
\end{theorem}

In fact, we will establish the following slightly stronger version of the theorem.  
It is stronger because it has an evidently weaker hypothesis:
\begin{theorem} \label{thm:mc2}
    Let $P$ be the transition operator for a Markov chain on~$\N$, and write $P_n = \avg_{t < n} \{P^t\}$.
    Suppose that $\trnorm{\wh{\pi} - \wh{\pi}P_n} \leq \eps < 1$ for all sufficiently large~$n$.
    Then there is an invariant distribution~$\pi$ for~$P$ with $\norm{\wh{\pi}- \pi}_1 \leq 3\eps$.
\end{theorem}

We begin with some notation:
\begin{notation}
    For $\sigma$ a vector indexed by~$\N$ and $S \subseteq \N$, we'll write $\sigma_S$ for the vector in which the entries of~$\sigma$ outside of~$S$ are zeroed out.
\end{notation}

As a key component of our proof, we will require the following known result from the theory of Markov chains on countably infinite state sets (immediate from, e.g.,~\cite[Cor.~2.1.4,~Thm.~3.2.3]{Por24}):
\begin{theorem}  \label{thm:class-decay}
    Let $P$ be the transition operator for a Markov chain on~$\N$.
    Write $\N = R_+ \sqcup D$, where $R_+$ is the set of positive recurrent states for~$P$, and $D = T \sqcup R_0$ is the set of transient and null recurrent states.
    Then for any distribution~$\pi$ on~$\N$ and any finite subset $S \subseteq D$, we have $(\pi P^t)_S \xrightarrow{t \to \infty} 0$.
\end{theorem}

We also use Scheff{\'e}'s lemma in our proof. Specifically, we use the following special case: 

\begin{lemma}[Scheff{\'e}'s lemma, special case]\label{lem:scheffe}
   Let $\{p_n\}$ be a sequence of probability distributions on a countable set $S$, and let $p$ be a probability distribution on $S$. Suppose that $p_n \xrightarrow{n\to\infty} p$ entrywise. Then $\norm{p_n - p}_1 \xrightarrow{n\to\infty} 0$.
\end{lemma}

We are now ready to establish our main result on Markov chains:
\begin{proof}[Proof of \Cref{thm:mc2}.]
    We first claim that \Cref{thm:class-decay} implies
    \begin{equation}    \label[ineq]{ineq:stuff}
        \norm{\wh{\pi}_D}_1 \leq \eps, \quad \text{which also implies } \norm{\wh{\pi}_{R_+}}_1 \geq 1-\eps > 0 \text{ and } \norm{\wh{\pi}_D P_n}_1 \leq \eps\ \forall t.
    \end{equation}
    For otherwise, select finite $S \subseteq D$ with $\norm{\wh{\pi}_{S}}_1 \geq \eps + \eta$ for some $\eta > 0$.
    By \Cref{thm:class-decay}, 
    we have $\norm{(\wh{\pi} P^t)_{S}}_1 \leq \eta/3$ for all sufficiently large~$t$, which implies $\norm{(\wh{\pi} P_n)_{S}}_1 \leq 2\eta/3$ for all sufficiently large~$n$.
    But our hypothesis on $\wh{\pi}$ is that 
    $\norm{\wh{\pi} - \wh{\pi}P_n}_1 \leq \eps$ for all sufficiently large~$n$, which implies that  $\norm{\wh{\pi}_S - (\wh{\pi}P_n)_S}_1 \leq \eps$ for all sufficiently large $n$.
    Hence, we must have 
    \[\norm{\wh{\pi}_{S}}_1 \leq \norm{\wh{\pi}_S - (\wh{\pi} P_n)_S}_1 + \norm{(\wh{\pi}P_n)_S}_1 \leq \eps + 2\eta/3,\] 
    a contradiction.

    Next, observe that for all sufficiently large $n \in \N$,
    \begin{equation}    \label[ineq]{ineq:above}
        \norm{\wh{\pi} - \wh{\pi}_{R_+} P_n}_1 
        \leq
        \norm{\wh{\pi} - \wh{\pi} P_n}_1 
        + 
        \norm{\wh{\pi}_D P_n}_1 \leq 2\eps
    \end{equation}
    by our hypothesis on~$\wh{\pi}$ and \Cref{ineq:stuff}.
    If we now write $\pi' = \wh{\pi}_{R_+}/\norm{\wh{\pi}_{R_+}}_1$, then $\pi'$ is a probability distribution supported on~$R_+$ satisfying $\norm{\pi' - \wh{\pi}_{R_+}}_1 \leq \eps$.
    Combining this with \Cref{ineq:above} and using contractivity of~$P_n$, we conclude
    \begin{equation}    \label[ineq]{ineq:3e}
        \norm{\wh{\pi} - \pi' P_n}_1 \leq 3\eps \text{ for all sufficiently large $n$.}
    \end{equation}
    Suppose for a moment that $P$ is irreducible when restricted to~$R_+$ (i.e., the digraph on~$R_+$ induced by~$P$'s positive entries is strongly connected).
    Then it is known~\cite[Thms.~2.2.8,~3.1.3]{Por24} that~$P$ has a unique invariant distribution~$\pi_0$, and moreover $\pi'$ (being supported on~$R_+$) satisfies
    \begin{equation}
        \pi' P_n \xrightarrow{n \to \infty} \pi_0 \text{ entrywise.}
    \end{equation}
    More generally, $R_+$ may be partitioned as $R_1 \sqcup R_2 \sqcup \cdots$ for some (possibly infinite) sequence such that $P$ is irreducible and positive-recurrent when restricted to each~$R_i$.  
    Then $\pi'$ may be written as a convex combination $\lambda_1 \pi'_1 + \lambda_2 \pi'_2 + \cdots$ of probability distributions, with $\lambda_i$ supported on~$R_i$; and, $P$ has a unique invariant distribution~$\pi_i$ on~$R_i$.  
    We then have that $\pi = \lambda_1 \pi_1 + \lambda_2\pi_2 + \cdots$ is an invariant distribution for~$P$, and that
    \begin{equation}
        \pi' P_n \xrightarrow{n \to \infty} \pi \text{ entrywise}.
    \end{equation}
    But $\norm{\pi' P_n}_1 = 1 = \norm{\pi}_1$ for all~$n$, so Scheff{\'e}'s lemma (\Cref{lem:scheffe}) implies $\norm{\pi' P_n - \pi}_1 \to 0$.
    We may now conclude $\norm{\wh{\pi}- \pi}_1 \leq 3\eps$ from \Cref{ineq:3e}.
\end{proof}

\section{Our upper bound for quantum CTC-assisted computation}

We now wish to prove  \Cref{thm:main-quantum}, that computation assisted by qubit-carrying CTCs of unbounded width and length still corresponds to~$\Delta_2$.
As discussed, the lower bound already follows from the classical case. 
To show that $\Delta_2$ remains an upper bound, almost everything is the same as in the classical upper bound in \Cref{sec:class-upper}. 
The assumption about computable quantum channels made at the end of \Cref{sec:infinite} takes the place of \Cref{lem:comp}.
We also make the analogous definition of everlasting-invariance:
\begin{definition}
    If $\Psi$ is a quantum channel, we say that mixed state $\wh{\rho}$ is \emph{$\eps$-everlasting-invariant} provided $\trnorm{\wh{\rho} - \Psi^t(\wh{\rho})}\leq \eps$ for all $t \in \N$.
\end{definition}
It is then straightforward to verify that the only task remaining to prove \Cref{thm:main-quantum} is establishing that, in the context of quantum channels, $\eps$-everlasting-invariant states are $\eps'$-close to truly invariant states (where $\eps' \to 0$ as $\eps \to 0$).
We do precisely this in the next section --- even (as in \Cref{thm:mc2}) with the weaker hypothesis that $\trnorm{\wh{\rho} - \Psi_n(\wh{\rho})}\leq \eps$, where $\Psi_n = \avg_{t < n}\{\Psi^t\}$.

\subsection{Our main quantum channel result}

We first establish an analogue of \Cref{thm:class-decay}.
(We remark that a continuous-time analogue also holds via the same proof.)

Before we begin, we remark that the remainder of this section uses concepts from Banach space theory and recently developed tools in the theory of quantum Markov maps. Preliminaries on the former can be found in textbooks (see, e.g., \cite[Appendix B]{Lan17} for one focused on connections to quantum mechanics). 
We refer readers to \cite{Gir22} and references therein for further detail on quantum Markov maps.

\begin{theorem} \label{thm:decay}
    Let $\Psi : \calB_1(\calH) \to \calB_1(\calH)$ be a quantum channel on a separable Hilbert space~$\calH$, and define $\Psi_n = \avg_{t < n}\{\Psi^t\}$.
    Write $\calH = \calD \oplus \calR_+$, where $\calR_+$ is the positive recurrent subspace and $\calD = \calT \oplus \calR_0$ is its orthogonal complement, i.e., the transient and null recurrent subspaces.
    Then for any state~$\rho$ and any projector~$\Pi$ onto a finite-dimensional subpsace of~$\calD$, we have $\tr(\Psi_n(\rho) \Pi) \xrightarrow{n\to\infty} 0$.
\end{theorem}
\begin{proof}
    The key result we use is due to Girotti~\cite[Thm.~2.3.23]{Gir22} (building on~\cite{Wat79,FV82,Luc95,Luc98,AGG02,CG21} et al.) which says that for every state~$\rho$, the following subnormalized state
    \begin{equation}   
        \calE_{\textrm{normal}\ast}(\rho) \coloneqq w^*\textrm{--}\lim_{\mathclap{n \to \infty}} \Psi_n(\rho)
    \end{equation}
    exists, and is supported on~$\calR_+$.
    The latter part of this assertion is because \cite[Thm.~2.3.23]{Gir22}
    shows $\calE_{\textrm{normal}\ast}$ is the predual of $\calE_{\textrm{normal}}$, the unique $w^*\text{-}w^*$-continuous ergodic projection for the quantum Markov semigroup given by powers of $\Psi^*$.  Then any operator in the range of $\calE_{\textrm{normal}\ast}$ is $\Psi$-invariant, and hence supported on~$\calR_+$. See \cite[Thm.~2.3.19]{Gir22} and the discussion immediately following the proof of \cite[Thm.~2.3.23]{Gir22}.

    Since $\calE_{\textrm{normal}\ast}(\rho)$ is supported on $\calR_+$, for any projector $\Pi$ onto a finite-dimensional subspace of $\calD = \calR_+^\bot$ we have
    \begin{equation}
    \tr(\calE_{\textrm{normal}\ast}(\rho)\Pi) = 0 \quad\implies\quad \tr(\Psi_n(\rho) \Pi) \xrightarrow{n\to\infty} 0
    \end{equation}
    because $\Pi$ is compact.
\end{proof}

We now give the main technical result in our paper:
\begin{theorem} \label{thm:channel}
    In the setting of \Cref{thm:decay}, suppose there exists a state $\wh{\rho}$ satisfying $\trnorm{\Psi_n(\wh{\rho})- \wh{\rho}} \leq \eps < 1$ for all sufficiently large~$n$.
    Then there is an invariant state~$\rho_\infty$ for~$\Psi$ with $\trnorm{\wh{\rho}- \rho_\infty} \leq 6\sqrt{\eps} + \eps$.
\end{theorem}
\begin{proof}
    Let $\Pi_{\calD}$ be the orthogonal projector onto~$\calD$.
    Now for any projector $\Pi \leq \Pi_{\calD}$ as in \Cref{thm:decay}, we have $\tr(\Psi_n(\wh{\rho}) \Pi) \to 0$.
    Our hypothesis on $\wh{\rho}$ implies that $\abs{\tr(\Psi_n(\wh{\rho}) \Pi) - \tr(\wh{\rho} \Pi)} \leq \eps$ for all sufficiently large~$n$, and so it follows that $\tr(\wh{\rho}\Pi) \leq \eps$.
    Since this holds for all $\Pi \leq \Pi_{\calD}$ with finite-dimensional range, we conclude that in fact 
    \begin{equation} \label[ineq]{ineq:qtrace}
        \tr(\wh{\rho}\Pi_{\calD}) \leq \eps \quad\implies\quad \tr(\wh{\rho}\Pi_{\calR_+}) \geq 1-\eps > 0,
    \end{equation}
    where $\Pi_{\calR_+}$ is the projector onto~$\calR_+$.
    We now define
    \begin{equation} 
        \rho_+ = \frac{\Pi_{\calR_+} \cdot \wh{\rho}  \cdot \Pi_{\calR_+}}{\tr(\wh{\rho} \Pi_{\calR_+})},
    \end{equation} 
    a normalized state, supported on $\calR_+$.
    It is known~\cite[Lem.~9]{Win99} that \Cref{ineq:qtrace} implies $\trnorm{\wh{\rho} - \rho_+} \leq \eps' \coloneqq 2\sqrt{\eps}$.
    Since $\Psi_n$ is a channel, it contracts in trace norm; thus
    also $\trnorm{\Psi_n(\wh{\rho}) - \Psi_n(\rho_+)} \leq \eps'$ for any~$n$, so $\trnorm{\wh{\rho} - \Psi_n(\rho_+)} \leq \eps' + \eps$ for sufficiently large~$n$, and we may conclude 
    \begin{equation} \label[ineq]{ineq:thebd}
        \trnorm{\rho_+  - \Psi_n(\rho_+)} \leq 2\eps' +\eps \quad \text{for sufficiently large } n.
    \end{equation}
    
Our remaining work will only involve analyzing $\Psi_n(\rho_+)$, and since $\rho_+$ is supported on the enclosure~$\calR_+$~\cite[Proposition 5.1]{CP16b}, we may henceforth restrict attention to $\Psi$'s action on~$\calR_+$.
    Now referring to the work of~\cite{CG21} (in particular, its Theorem~19), the absorption operator~$\calA(\calR_+)$ for this restricted channel equals~$\Id$, and hence 
    \begin{equation}    \label{eq:rhoinf}
        \rho_{\infty} \coloneqq w\textrm{--}\lim_{\mathclap{n \to \infty}} \Psi_n(\rho_+)
    \end{equation}
    exists and is an invariant state for~$\Psi$.
Then by \Cref{ineq:thebd} and \Cref{eq:rhoinf} we have
    \begin{equation}
    \norm{\rho_+ - \rho_\infty}_1 \leq 2\eps' + \eps + \norm{\Psi_n(\rho_+) - \rho_\infty}_1 \xrightarrow{n \to \infty} 2 \eps' + \eps.
    \end{equation}
    The last step follows from the variational definition of $\norm{{\cdot}}_1$ combined with the fact that \[\tr(\Psi_n(\rho_+)X) \to \tr(\rho_\infty X)\] for all operators $X \in \calB(\calR_+)$ by \Cref{eq:rhoinf}.
    So we get $\trnorm{\rho_+ - \rho_\infty} \leq 2\eps' + \eps$ and thus $\trnorm{\wh{\rho} - \rho_\infty} \leq 3\eps' + \eps$, completing the proof.
\end{proof}

\section{Hardness results for classifying Markov chain states} \label{sec:hardness-classify}

In this section we show $\Sigma_2$- and $\Pi_2$-hardness results for classifying states in a (computable) Markov chain as transient/null recurrent/positive recurrent.
These results arguably make our main result --- that (approximate) invariant distributions for a given chain (or channel) can be found in the provably smaller class~$\Delta_2$ --- a bit more surprising.
For example, after the $\Delta_2$-algorithm has found a distribution~$\wh{\pi}$ satisfying $\trnorm{\wh{\pi} - \pi} \leq \eps$ for some invariant~$\pi$, it knows that this~$\pi$ is entirely supported on positive recurrent states, but it has no idea \emph{which} states in the support of~$\wh{\pi}$ are in the ``wrong'' classes (transient/null recurrent) and which are in the ``right'' one (positive recurrent).

\begin{theorem} \label{thm:posrecsig}
    Given as input (a TM $C$ computing) a Markov chain~$P$ on~$\{0,1\}^*$, as well as a state~$x \in \{0,1\}^*$, the problem of deciding whether~$x$ is a positive recurrent state is $\Sigma_2$-complete.
\end{theorem}
\begin{proof}
    We must prove containment and hardness.

    \paragraph{The problem is in~$\Sigma_2$.}
    For this, the idea is that $x$ is positive recurrent iff there is an invariant state for~$P$ with positive probability mass on~$x$.  
    Our $\Sigma_2$ algorithm will  first nondeterministically guess a rational $\eps > 0$ and a finitely supported rational probability vector~$\wh{\pi}$. Then it will use the~$\Delta_2$ subroutine implied by \Cref{thm:class-approx} to decide if~$\wh{\pi}$ is $\eps$-close to some invariant distribution for~$P$, or else $6\eps$-far from every invariant distribution.
    Finally, if the $\Delta_2$ subroutine accepts (meaning $\wh{\pi}$ is definitely within~$6\eps$ of an invariant distribution), our $\Sigma_2$ algorithm  will accept iff $\wh{\pi}_x \geq 7\eps$.

    To show correctness, first observe that if the overall algorithm accepts, then there must be an invariant distribution~$\pi$ with $\trnorm{\wh{\pi} - \pi} \leq 6\eps$, and then $\wh{\pi}_x \geq 7\eps$ implies $\pi_x \geq \eps > 0$, so $x$ is indeed positive recurrent.
    On the other hand, if $x$ is positive recurrent, then there is some invariant~$\pi$ and some rational~$\delta$ with $\pi_x \geq \delta > 0$.  
    Now when our $\Sigma_2$ algorithm guesses $\eps = \delta/8$ and some $\wh{\pi}$ satisfying $\trnorm{\wh{\pi} - \pi} \leq \eps$, the following will hold true: $\wh{\pi}_x \geq \pi_x - \eps \geq (7/8)\delta$; the $\Delta_2$-subroutine will accept~$\wh{\pi}$; and, the $\Sigma_2$ algorithm will verify $\wh{\pi}_x \geq 7\eps = (7/8)\delta$.
    Thus the $\Sigma_2$ algorithm will accept all positive recurrent~$x$.

    \paragraph{$\Sigma_2$-hardness.}
    In \cite{KKM19} it was shown that the following task is $\Sigma_2$-hard:  Given a probabilistic Turing Machine~$M$ and an input~$x$, decide if $\E[\text{running time of }M(x)] < \infty$.  
    We give a polynomial-time reduction from this task to that of deciding positive recurrence.
    So suppose the reduction is given~$M$ and~$x$ as input.
    The reduction will first slightly modify~$M'$ so that it has the property that it can never re-enter its initial configuration.\footnote{This can be done without changing the running time of $M$ by  having~$M'$ mark the initial tape square with a special symbol that is subsequently ignored.}
    Now~$M'$ can essentially be regarded as a Markov chain, with state space equal to all its possible TM-configurations.  
    Let $x'$ denote the state encoding the initial TM-configuration with input~$x$.
    Our reduction will output~$x'$ together with (a Turing Machine computing) the Markov chain~$P$ corresponding to~$M'$ --- with the twist that in~$P$, all halting configuration states transition back to state~$x'$ with probability~$1$.

    This $P$ has the property that when started at state~$x'$, it simulates the computation of~$M'$ on~$x$.
    Moreover, the chain re-enters~$x'$ iff the simulation of~$M(x)$ halts.  
    Thus $x'$ is positive recurrent for~$P$ iff the expected number of steps to return to~$x'$ when starting from~$x'$ is finite iff the expected running time of~$M(x)$ is finite.
\end{proof}

\begin{theorem} \label{thm:transientsig}
    Given as input (a TM $C$ computing) a Markov chain~$P$ on~$\{0,1\}^*$, as well as a state~$x \in \{0,1\}^*$, the problem of deciding whether~$x$ is a transient state is $\Sigma_2$-complete.
\end{theorem}
\begin{proof}
    Again we prove containment and hardness.

    \paragraph{The problem is in~$\Sigma_2$.}   
    A $\Sigma_2$ algorithm for verifying that $x$ is transient for~$P$ is as follows.
    First, existentially guess a rational $\eps > 0$.
    Then, universally guess a time~$t \in \N$.
    Finally, use the techniques of \Cref{lem:comp} to compute an estimate $\wh{p}_t$ that is within $\eps$ of
    \begin{equation}
        p_t \coloneqq \Pr[P \text{ visits } x \text{ within $t$ steps when starting from $x$}].
    \end{equation}
    Finally, accept iff $\wh{p}_t \leq 1-2\eps$.

    To verify correctness, first suppose the algorithm accepts.  
    Then the algorithm has verified that $p_t \leq 1-\eps$ for all~$t$, which means that the probability~$P$ \emph{ever} returns to $x$ when starting from~$x$ is at most $1-\eps$; hence $x$ is a transient state.
    On the other hand, if $x$ is transient then there exists rational $\delta > 0$ such that $p_t \leq 1-\delta$ for all $t$; thus when the algorithm guesses $\eps = \delta/2$, it will end up accepting.

    \paragraph{$\Sigma_2$-hardness.}
    In \cite{KKM19} it was shown that the following task is $\Sigma_2$-hard:  Given a probabilistic Turing Machine~$M$ and an input~$x$, decide if $\Pr[M(x) \text{ halts}] < 1$.
    To reduce from this task to that of deciding transience, we use the exact same reduction as in the proof of \Cref{thm:posrecsig}.
    Then the probability of $P$ returning to state~$x'$ when starting from~$x'$ is the same as the probability of~$M(x)$ halting; this establishes correctness of the reduction.
\end{proof}

\begin{theorem}
    Given as input (a TM $C$ computing) a Markov chain~$P$ on~$\{0,1\}^*$, as well as a state~$x \in \{0,1\}^*$, the problem of deciding whether~$x$ is a null recurrent state is $\Pi_2$-complete.
\end{theorem}
\begin{proof}
    This is immediate from \Cref{thm:posrecsig,thm:transientsig}, since every state is either transient, null recurrent, or positive recurrent.
\end{proof}

\section{Postselected CTCs} \label{sec:PCTC}
\emph{Postselected (teleportation) CTCs} (henceforth ``P-CTCs'') are an alternative to the Deutschian model of CTCs (``D-CTCs'') that we have studied so far.
They were introduced independently by Bennett and Schumacher~\cite{BS05} and by Svetlichny~\cite{Sve09,Sve11}, and were developed further in e.g.~\cite{LMG+11,LMGb+11,BW12}.
One way of describing P-CTCs is via the quantum teleportation protocol:  After Alice and Bob share an EPR pair to facilitate teleportation, Bob can begin using his qubit as though he had already received Alice's teleported state $\ket{\psi}$ from the future (before Alice has even decided on it).  When Alice later completes the last local step the protocol (namely, measuring in the Bell basis), on the $1/4$-probability chance that her readout is $\ket{\Phi^+}$, Bob need not do anything to (have) possess(ed)~$\ket{\psi}$.  Thus we have a ``time travel''-like phenomenon\dots subject to ``postselecting'' on a particular measurement outcome for Alice.
(Indeed, \cite{BS05} described a purely classical version of the P-CTC model in which postselection ``can be used to simulate time travel without the need of any exotic equipment''.\footnote{``\emph{Q: Is it time travel?  A: It depends on what your definition of `is' is.}'' --- Charles Bennett~\cite{BS05}})  As described by, e.g., Brun and Wilde~\cite{BW12}, having a ``P-CTC'' is equivalent to allowing for postselection on one element~$M_1$ of a general quantum measurement $(M_1, \dots, M_k)$ (subject to the constraint that the probablity of~$M_1$ must be nonzero).  
Thus from the perspective of computational complexity, it is equivalent to adding the postselection operator $\post$ discussed towards the end of \Cref{sec:prior}.
Indeed, Lloyd et al.~\cite{LMG+11,LMGb+11,BW12} noted that in the context of polynomial-time (bounded error) computation, P-CTCs grant the power of~$\PP$.

In the spirit of the present paper, we may ask what is the computational power of P-CTCs in the context of unbounded-time probabilistic computations with unbounded error probability.  
It turns out it is the same as it is without P-CTCs, and even without randomness:
\begin{proposition}\label{prop:postR-R}
    Suppose $L$ is computable with postselection and with unbounded error; that is, there is a probabilistic TM $M$ with the following properties:
    \begin{itemize}
        \item On every input~$x$ we have that $M(x)$ halts with probability~$1$ and outputs either $0$, $1$, or $?$.\footnote{One should think of postselecting on the event that $M(x) \neq\ ?$.}
        \item $x \in L \implies \Pr[M(x) = 1] > \Pr[M(x) = 0]$.
        \item $x \not \in L \implies \Pr[M(x) = 0] > \Pr[M(x) = 1]$.
    \end{itemize}
    Then $L \in \Rec$; i.e., $L$ is decidable.
\end{proposition}

This result has a simple proof using the theory of c.e.~reals (aka computably enumerable reals, or left-semicomputable reals), which dates back to Soare~\cite{Soa67}; see, e.g.,~\cite{CHKW98,AWZ00,DWZ04}.
A real number in~$[0,1]$ is c.e.~if and only if it is the acceptance probability of some probabilistic Turing Machine (on some input); in general, $x \in \R$ is c.e.\ if and only if $x$~mod~$1$ is.  
\begin{proof}
    Let $p^x_0$ (respectively, $p^x_1$, $p^x_?$) be the probability that $M(x)$ outputs~$0$ (respectively, $1$, $?$).
    We have that $p^x_0, p^x_1,p^x_?$ are c.e.\ and satisfy $p^x_0 + p^x_1+p^x_? = 1$; moreover
    \begin{align}
        x \in L &\iff p^x_1 > p^x_0 = 1 - p^x_1 - p^x_? \iff q^x_1 \coloneqq \tfrac23 p^x_1 + \tfrac13 p^x_?  > \tfrac13 \label{eqn:aa},\\
        x \not \in L &\iff p^x_0 > p^x_1 = 1 - p^x_0 - p^x_? \iff q^x_0 \coloneqq \tfrac23 p^x_0 + \tfrac13 p^x_?  > \tfrac13. \label{eqn:bb}
    \end{align}
    But finite convex combinations of c.e.~probabilities are c.e., and in a uniform way: for $i \in \{0,1\}$, there is a probabilistic TM $M_i$ that halts on~$x$ with probability~$q^x_i$.
    Moreover, for any rational like~$\frac13$ we have that $q^x_i > \frac13$ is semidecidable; so, to decide $x \stackrel{?}{\in} L$, it suffices to check which of~\Cref{eqn:aa,eqn:bb} holds.
    More precisely, the algorithm for deciding $x \stackrel{?}{\in} L$ approximates $\Pr[M_0(x) \text{ halts}]$ and $\Pr[M_1(x) \text{ halts}]$ from below in parallel, and outputs $i$ as soon as its approximation for $\Pr[M_i(x) \text{ halts}] = q^x_i$ exceeds~$\frac13$.
\end{proof}

We might also consider relaxing the condition in \Cref{prop:postR-R} that $M(x)$ halts with probability~$1$; perhaps this might correspond to P-CTC computations that need not terminate.  
Let us first characterize the class of languages decided with \emph{bounded} $2$-sided error using such P-CTCs:
\begin{proposition} \label{prop:u2}
    Let $\calC$ be the class of languages~$L$ computable in the following sense: there is a probabilistic TM $M$ with the following properties:\footnote{The interpretation is that, postselecting on the event that $M(x)$ halts, it outputs the correct decision with probability $> \frac23$.}
    \begin{itemize}
        \item $x \in L \implies \Pr[M(x) = 1] > 2\Pr[M(x) = 0]$.
        \item $x \not \in L \implies \Pr[M(x) = 0] > 2\Pr[M(x) = 1]$.
    \end{itemize}
    Then $\calC = \Rec_{\|}^{\Halts}$.
\end{proposition}
\begin{proof}
    ($\calC \subseteq \Rec_{\|}^{\Halts}$): Suppose $L \in \calC$ via TM~$M$. 
    Then a TM $M'$ can decided $L$ by truth-table reduction to $\Halts$ as follows:  For any input~$x$, let us write (using notation from  the previous proof) $\ul{p}^x \coloneqq \min(p^x_0, p^x_1)$, $\ol{p}^x \coloneqq \max(p^x_0, p^x_1)$, and $r^x \coloneqq \frac{p^x_0 + p^x_1}{2}$.
    Since $r^x$ is c.e.\ and positive by assumption, $M'$ can compute some rational $\eps > 0$ such that $r^x > \eps$, and hence also $\ol{p}^x > \eps$.
    But now since $\ul{p}^x$ and $\ol{p}^x$ differ by a factor of at least~$2$, it follows that (at least) one of the finitely many numbers $\theta \in T \coloneqq \{\eps, 2\eps, 4\eps, 8\eps, \dots\} \cap (0,1)$ must satisfy $\ul{p}^x  \leq \theta < \ol{p}^x$.
    It is easy to see that the $\Halts$ oracle can be used to decide $p_i^x  \leq \theta$ and $\theta < p_i^x$ for any~$\theta$ and for both $i \in \{0,1\}$.  
    So by making $2|T|$ nonadaptive queries to the oracle, $M'$ can find such a~$\theta$ that splits $p_0^x$ and $p_1^x$, and thereby output the $j \in \{0,1\}$ for which $p_j^x$ is larger.

\bigskip

    ($\Rec_{\|}^{\Halts} \subseteq \calC$): Suppose $L \in \Rec_{\|}^{\Halts}$, so $L$ is decided by some TM $N$ that makes nonadaptive calls to a $\Halts$ oracle.  We now define a probabilistic TM $M$ that has the two required properties in the proposition. 
    (And in fact, we can arrange for, e.g., factor ``$99$'' in place of factor~``$2$''; i.e., the postselected $M$ has success probability exceeding~$.99$.)
    $M(x)$ will begin by simulating $N(x)$ to produce its oracle calls, $B_1, \dots, B_k$.  
    Let $b_1, \dots, b_k \in \{0,1\}$ denote the correct oracle responses (meaning $b_i = 1$ iff $B_i$ halts on the empty tape).  
    Of course, $M(x)$ does not know these; it will instead produce independent ``guesses'' $\bg_1, \dots, \bg_k$ for each, where $\bg_i$ is chosen to be~$0$ with probability $\lambda \coloneqq \frac{1}{99k}$, and~$1$ with probability $1-\lambda$.
    $N(x)$ will then attempt to certify its $1$-guesses; i.e., in parallel for each $i$ with $\bg_i = 1$, the algorithm will simulate~$B_i$ to try to check that it indeed halts on empty input.  
    Whenever $M(x)$ has at least one wrong guess $\bg_i = 1$ (meaning $b_i = 0$, i.e., $B_i$ does not halt), then $M(x)$ will never halt.  
    Finally, in case $M(x)$ certifies all its $1$-guesses, it outputs whatever $N(x)$ would output had it received oracle responses $\bg_1, \dots, \bg_k$.

    We see that $M(x)$ will halt iff $(\bg_1, \dots, \bg_k) \leq (b_1, \dots, b_k)$ entrywise. 
    This occurs with probability exactly $\lambda^{k-|b|}$.
    Moreover, $M(x)$ will halt with the \emph{correct} answer whenever $(\bg_1, \dots, \bg_k) = (b_1, \dots, b_k)$; conditioned on halting, this occurs with probability $(1-\lambda)^{|b|} \geq (1-\lambda)^k > 1-\lambda k \geq .99$.
    This completes the proof.
\end{proof}

It is also a natural question to characterize the complexity class that results when one uses \emph{unbounded} error in \Cref{prop:u2}; i.e., when its conditions are changed to $x \in L \implies \Pr[M(x) = 1] > \Pr[M(x) = 0]$ and $x \not \in L \implies \Pr[M(x) = 0] > \Pr[M(x) = 1]$.
We leave this open.

\section*{Acknowledgments}
The authors of \cite{ABG16} thank Matt Coudron, Michael Devin, Matt Hastings, Michael Nielsen, and Cem Say for helpful discussions. We also thank the Park City Math Institute for a summer school that brought some of the authors together.

\bibliographystyle{alpha}
\bibliography{odonnell}

\newcommand{\etalchar}[1]{$^{#1}$}
\begin{thebibliography}{LMGP{\etalchar{+}}11b}

\bibitem[Aar04a]{Aar04a}
Scott Aaronson.
\newblock Is quantum mechanics an island in {T}heoryspace?
\newblock Technical Report quant-ph/0401062, arXiv, 2004.

\bibitem[Aar04b]{Aar04b}
Scott Aaronson.
\newblock {\em Limits on Efficient Computation in the Physical World}.
\newblock PhD thesis, University of California, Berkeley, 2004.

\bibitem[Aar05a]{Aar05}
Scott Aaronson.
\newblock {NP}-complete problems and physical reality.
\newblock {\em ACM SIGACT News}, 36(1):30--52, 2005.

\bibitem[Aar05b]{Aar05b}
Scott Aaronson.
\newblock Quantum computing, postselection, and probabilistic polynomial-time.
\newblock {\em Proceedings of the Royal Society A}, 461(2063):3473--3482, 2005.

\bibitem[ABG16]{ABG16}
Scott Aaronson, Mohammad Bavarian, and Giulio Gueltrini.
\newblock Computability theory of closed timelike curves.
\newblock Technical Report quant-ph/1609.05507, arXiv, 2016.

\bibitem[AFF{\etalchar{+}}01]{AFF+01}
James Aspnes, David Fischer, Michael Fischer, Ming-Yang Kao, and Alok Kumar.
\newblock Towards understanding the predictability of stock markets from the perspective of computational complexity.
\newblock In {\em Proceedings of the 12th Annual ACM-SIAM Symposium on Discrete Algorithms}, pages 745--754, 2001.

\bibitem[AGG02]{AGG02}
Alvaro Arias, Aurelian Gheondea, and Stanley Gudder.
\newblock Fixed points of quantum operations.
\newblock {\em Journal of Mathematical Physics}, 43(12):5872--5881, 2002.

\bibitem[AW09]{AW09}
Scott Aaronson and John Watrous.
\newblock Closed timelike curves make quantum and classical computing equivalent.
\newblock {\em Proceedings of the Royal Society A}, 465(2102):631--647, 2009.

\bibitem[AWZ00]{AWZ00}
Klaus Ambos{-}Spies, Klaus Weihrauch, and Xizhong Zheng.
\newblock Weakly computable real numbers.
\newblock {\em Journal of Complexity}, 16(4):676--690, 2000.

\bibitem[Bac04]{Bac04}
Dave Bacon.
\newblock Quantum computational complexity in the presence of closed timelike curves.
\newblock {\em Physical Review A}, 70(3):032309, 2004.

\bibitem[BG05]{BG05}
Olivier Bournez and Florent Garnier.
\newblock Proving positive almost-sure termination.
\newblock In {\em International Conference on Rewriting Techniques and Applications}, pages 323--337. Springer, 2005.

\bibitem[BGM03]{BGM03a}
Elmar B{\"o}hler, Christian Gla{\ss}er, and Daniel Meister.
\newblock Error-bounded probabilistic computations between {MA} and {AM}.
\newblock In {\em Proceedings of the 28th Annual International Symposium on Mathematical Foundations of Computer Science}, pages 249--258, 2003.

\bibitem[BN12]{BN12}
Bernhard Baumgartner and Heide Narnhofer.
\newblock The structures of state space concerning quantum dynamical semigroups.
\newblock {\em Reviews in Mathematical Physics}, 24(2):1250001, 30, 2012.

\bibitem[Bru03]{Bru03}
Todd Brun.
\newblock Computers with closed timelike curves can solve hard problems efficiently.
\newblock {\em Foundations of Physics Letters}, 16(3):245--253, 2003.

\bibitem[BS05]{BS05}
Charles Bennett and Benjamin Schumacher.
\newblock Teleportation, simulated time travel, and how to flirt with someone who has fallen into a black hole.
\newblock Talk in QUPON. \url{http://web.archive.org/web/20060207123032/http://www.research.ibm.com/people/b/bennetc/QUPONBshort.pdf}, 2005.

\bibitem[BW12]{BW12}
Todd Brun and Mark Wilde.
\newblock Perfect state distinguishability and computational speedups with postselected closed timelike curves.
\newblock {\em Foundations of Physics}, 42(3):341--361, 2012.

\bibitem[Car77]{Car72}
Hans~Georg Carstens.
\newblock {$\Delta \sp{0}\sb{2}$}-{M}engen.
\newblock {\em Archiv f\"{u}r Mathematische Logik und Grundlagenforschung}, 18(1-2):55--65, 1976/77.

\bibitem[CG21]{CG21}
Raffaella Carbone and Federico Girotti.
\newblock Absorption in invariant domains for semigroups of quantum channels.
\newblock {\em Annales Henri Poincar\'{e}}, 22(8):2497--2530, 2021.

\bibitem[Cha75]{Cha75}
Gregory Chaitin.
\newblock A theory of program size formally identical to information theory.
\newblock {\em Journal of the Association for Computing Machinery}, 22:329--340, 1975.

\bibitem[CHKW98]{CHKW98}
Cristian Calude, Peter Hertling, Bakhadyr Khoussainov, and Yongge Wang.
\newblock Recursively enumerable reals and {C}haitin {$\Omega$} numbers.
\newblock In {\em Proceedings of the 5th Annual Symposium on Theoretical Aspects of Computer Science}, pages 596--606, 1998.

\bibitem[CP16a]{CP16b}
Raffaella Carbone and Yan Pautrat.
\newblock Irreducible decompositions and stationary states of quantum channels.
\newblock {\em Reports on Mathematical Physics}, 77(3):293--313, 2016.

\bibitem[CP16b]{CP16}
Raffaella Carbone and Yan Pautrat.
\newblock Open quantum random walks: reducibility, period, ergodic properties.
\newblock {\em Annales Henri Poincar\'{e}}, 17(1):99--135, 2016.

\bibitem[CS13]{CS13}
Aleksandar Chakarov and Sriram Sankaranarayanan.
\newblock Probabilistic program analysis with martingales.
\newblock In {\em Computer Aided Verification}, pages 511--526. Springer, 2013.

\bibitem[Deu91]{Deu91}
David Deutsch.
\newblock Quantum mechanics near closed timelike lines.
\newblock {\em Physical Review D}, 44(10):3197--3217, 1991.

\bibitem[DWZ04]{DWZ04}
Rod Downey, Guohua Wu, and Xizhong Zheng.
\newblock Degrees of d.c.e.\ reals.
\newblock {\em Mathematical Logic Quarterly}, 50(4-5):345--350, 2004.

\bibitem[EGK12]{EGK12}
Javier Esparza, Andreas Gaiser, and Stefan Kiefer.
\newblock Proving termination of probabilistic programs using patterns.
\newblock In {\em Computer Aided Verification}, pages 123--138. Springer, 2012.

\bibitem[Ers68]{Ers68}
Yurii Ershov.
\newblock A certain hierarchy of sets. {I}.
\newblock {\em Algebra i Logika}, 7(1):47--74, 1968.

\bibitem[Fag04]{Fag04}
Franco Fagnola.
\newblock Quantum {M}arkov semigroups: structure and asymptotics.
\newblock {\em Rendiconti del Circolo Matematico di Palermo. Serie II.}, 73:35--51, 2004.

\bibitem[FR03]{FR03}
Franco Fagnola and Rolando Rebolledo.
\newblock Transience and recurrence of quantum {M}arkov semigroups.
\newblock {\em Probability Theory and Related Fields}, 126(2):289--306, 2003.

\bibitem[FV82]{FV82}
Alberto Frigerio and Maurizio Verri.
\newblock Long-time asymptotic properties of dynamical semigroups on {$W\sp{\ast} $}-algebras.
\newblock {\em Mathematische Zeitschrift}, 180(2):275--286, 1982.

\bibitem[Gir22]{Gir22}
Federico Girotti.
\newblock {\em Absorption in invariant domains for quantum {M}arkov evolutions}.
\newblock PhD thesis, Universit{\`a} degli Studi di Pavia, 2022.

\bibitem[GK12]{GK12}
Andreas G{\"a}rtner and Burkhard K{\"u}mmerer.
\newblock A coherent approach to recurrence and transience for quantum {M}arkov operators.
\newblock Technical Report 1211.6876, arXiv, 2012.

\bibitem[G{\"o}d49a]{Goe49}
Kurt G{\"o}del.
\newblock An example of a new type of cosmological solutions of {E}instein's field equations of gravitation.
\newblock {\em Reviews of Modern Physics}, 21(3):447, 1949.

\bibitem[G{\"o}d49b]{God49}
Kurt G{\"o}del.
\newblock Remark about the relationship between relativity theory and idealistic philosophy.
\newblock In {\em The Library of Living Philosophers, Volume 7. Albert Einstein: Philosopher-Scientist}, pages 557--562. Open Court, 1949.

\bibitem[HE73]{HE73}
Stephen Hawking and George Ellis.
\newblock {\em The large scale structure of space-time}.
\newblock Cambridge Monographs on Mathematical Physics, No. 1. Cambridge University Press, London-New York, 1973.

\bibitem[HHT93]{HHT93}
Yenjo Hem, Lane Hemaspaandra, and Thomas Thierauf.
\newblock Threshold computation and cryptographic security.
\newblock In {\em Proceedings of the 4th Annual International Symposium on Algorithms and Computation}, pages 230--239, 1993.

\bibitem[HHT97]{HHT97}
Yenjo Han, Lane Hemaspaandra, and Thomas Thierauf.
\newblock Threshold computation and cryptographic security.
\newblock {\em SIAM Journal on Computing}, 26(1):59--78, 1997.

\bibitem[HSP83]{HSP83}
Sergiu Hart, Micha Sharir, and Amir Pnueli.
\newblock Termination of probabilistic concurrent program.
\newblock {\em ACM Transactions on Programming Languages and Systems (TOPLAS)}, 5(3):356--380, 1983.

\bibitem[KKM19]{KKM19}
Benjamin Kaminski, Joost-Pieter Katoen, and Christoph Matheja.
\newblock On the hardness of analyzing probabilistic programs.
\newblock {\em Acta Informatica}, 56:255--285, 2019.

\bibitem[KKMO16]{KKMO16}
Benjamin Kaminski, Joost-Pieter Katoen, Christoph Matheja, and Federico Olmedo.
\newblock Weakest precondition reasoning for expected run--times of probabilistic programs.
\newblock In {\em Programming Languages and Systems}, pages 364--389. Springer, 2016.

\bibitem[Koz79]{Koz79}
Dexter Kozen.
\newblock Semantics of probabilistic programs.
\newblock In {\em Proceedings of the 20th Annual IEEE Symposium on Foundations of Computer Science}, pages 101--114. IEEE, New York, 1979.

\bibitem[Lan24]{Lan24}
Kornel Lanczos.
\newblock {\"U}ber eine station{\"a}re kosmologie im {S}inne der {E}insteinschen {G}ravitationstheorie.
\newblock {\em Zeitschrift f{\"u}r Physik}, 21(1):73--110, 1924.

\bibitem[Lan17]{Lan17}
Klaas Landsman.
\newblock {\em Foundations of quantum theory: From classical concepts to operator algebras}.
\newblock Springer Nature, 2017.

\bibitem[LMGP{\etalchar{+}}11a]{LMG+11}
Seth Lloyd, Lorenzo Maccone, Ra\'{u}l Garc\'{i}a-Patr\'{o}n, Vittorio Giovannetti, and Yutaka Shikano.
\newblock Quantum mechanics of time travel through post-selected teleportation.
\newblock {\em Physical Review D}, 84(2):025007, 2011.

\bibitem[LMGP{\etalchar{+}}11b]{LMGb+11}
Seth Lloyd, Lorenzo Maccone, Ra\'{u}l Garc\'{i}a-Patr\'{o}n, Vittorio Giovannetti, Yutaka Shikano, Stefano Pirandola, Lee Rozema, Ardavan Darabi, Yasaman Soudagar, Lynden Shalm, and Aephraim Steinberg.
\newblock Closed timelike curves via postselection: theory and experimental test of consistency.
\newblock {\em Physical Review Letters}, 106(4):040403, 2011.

\bibitem[{\L}uc95]{Luc95}
Andrzej {\L}uczak.
\newblock Ergodic projection for quantum dynamical semigroups.
\newblock {\em International Journal of Theoretical Physics}, 34(8):1533--1540, 1995.

\bibitem[{\L}uc98]{Luc98}
Andrzej {\L}uczak.
\newblock Properties of ergodic projection for quantum dynamical semigroups.
\newblock {\em International Journal of Theoretical Physics}, 37(1):555--562, 1998.

\bibitem[Nor98]{Nor98}
James Norris.
\newblock {\em Markov Chains}.
\newblock Cambridge University Press, 1998.

\bibitem[OS14]{OS14}
Ryan O'Donnell and A.~C.~Cem Say.
\newblock One time-traveling bit is as good as logarithmically many.
\newblock In {\em Proceedings of the 35th Annual IARCS Conference on Foundations of Software Technology and Theoretical Computer Science}, 2014.

\bibitem[OS18]{OS18}
Ryan O'Donnell and A.~C.~Cem Say.
\newblock The weakness of {CTC} qubits and the power of approximate counting.
\newblock {\em ACM Transactions on Computation Theory}, 10(2):1--22, 2018.

\bibitem[Por24]{Por24}
Ursula Porod.
\newblock {\em Dynamics of {M}arkov Chains for Undergraduates}.
\newblock \url{https://www.math.northwestern.edu/documents/book-markov-chains.pdf}, 2024.
\newblock February 27 edition.

\bibitem[Raa23]{Raa23}
Marien Raat.
\newblock Computation in nested closed timelike curves.
\newblock Master's thesis, Utrecht University, 2023.
\newblock \url{https://studenttheses.uu.nl/bitstream/handle/20.500.12932/45273/thesis-marien-raat.pdf}.

\bibitem[Sim75]{Sim75}
Janos Simon.
\newblock {\em On some central problems in computational complexity}.
\newblock PhD thesis, Cornell University, 1975.
\newblock TR 75-224.

\bibitem[Soa67]{Soa67}
Robert Soare.
\newblock {\em Recursion theory and {D}edekind cuts}.
\newblock PhD thesis, Cornell University, 1967.

\bibitem[Soa87]{Soa87}
Robert Soare.
\newblock {\em Recursively Enumerable Sets and Degrees: A Study of Computable Functions and Computably Generated Sets}.
\newblock Springer Berlin, Heidelberg, 1987.

\bibitem[SPH84]{SPH84}
Micha Sharir, Amir Pnueli, and Sergiu Hart.
\newblock Verification of probabilistic programs.
\newblock {\em SIAM Journal on Computing}, 13(2):292--314, 1984.

\bibitem[SU06]{SU06}
Ronen Shaltiel and Christopher Umans.
\newblock Pseudorandomness for approximate counting and sampling.
\newblock {\em Computational Complexity}, 15(4):298--341, 2006.

\bibitem[Sve00]{Sve09}
George Svetlichny.
\newblock Effective quantum time travel.
\newblock Technical Report 0902.4898, arXiv, 2000.

\bibitem[Sve11]{Sve11}
George Svetlichny.
\newblock Time travel: {D}eutsch vs.\ teleportation.
\newblock {\em International Journal of Theoretical Physics}, 50(12):3903--3914, 2011.

\bibitem[SY12]{SY12}
A.~C.~Cem Say and Abuzer Yakary{\i}lmaz.
\newblock Computation with multiple {CTCs} of fixed length and width.
\newblock {\em Natural Computing}, 11(4):579--594, 2012.

\bibitem[Uma06]{Uma06}
Veronica Umanit\`a.
\newblock Classification and decomposition of quantum {M}arkov semigroups.
\newblock {\em Probability Theory and Related Fields}, 134(4):603--623, 2006.

\bibitem[vS38]{Sto38}
Willem van Stockum.
\newblock The gravitational field of a distribution of particles rotating about an axis of symmetry.
\newblock {\em Proceedings of the Royal Society of Edinburgh}, 57:135--154, 1938.

\bibitem[Wat79]{Wat79}
Seiji Watanabe.
\newblock Ergodic theorems for dynamical semi-groups on operator algebras.
\newblock {\em Hokkaido Mathematical Journal}, 8(2):176--190, 1979.

\bibitem[Wat99]{Wat99}
John Watrous.
\newblock Space-bounded quantum complexity.
\newblock {\em Journal of Computer and System Sciences}, 59(2):281--326, 1999.

\bibitem[Wat11]{Wat11}
John Watrous.
\newblock An introduction to quantum information and quantum circuits.
\newblock {\em ACM SIGACT News}, 42(2):52--67, 2011.

\bibitem[Win99]{Win99}
Andreas Winter.
\newblock Coding theorem and strong converse for quantum channels.
\newblock {\em IEEE Transactions on Information Theory}, 45(7):2481--2485, 1999.

\bibitem[Yin12]{Ying12}
Mingsheng Ying.
\newblock Floyd--{H}oare logic for quantum programs.
\newblock {\em ACM Transactions on Programming Languages and Systems}, 33(6):1--49, 2012.

\end{thebibliography}

\end{document}